\documentclass[12pt,a4paper]{article}
\usepackage{amsmath}
\usepackage{amssymb}
\usepackage{colordvi}
\usepackage{epsfig}
\usepackage{subfigure}
\usepackage{mathrsfs}
\usepackage{cite}

\newcommand{\eq}[1]{Eq.~(\ref{#1})}
\newcommand{\eqs}[1]{Eqs.~(\ref{#1})}
\newcommand{\fig}[1]{Figure\ \ref{#1}}

\newcommand{\Bd}{B_d^0}
\newcommand{\Bbd}{\bar B_d^0}
\newcommand{\Bp}{B^+}
\newcommand{\Bm}{B^-}

\newcommand{\Apm}{A_{+-}}
\newcommand{\ApmB}{\bar A_{+-}}
\newcommand{\Aoo}{A_{00}}
\newcommand{\AooB}{\bar A_{00}}
\newcommand{\Apo}{A_{+0}}
\newcommand{\ApoB}{\bar A_{+0}}

\newcommand{\Bpm}{B^{+-}}
\newcommand{\Bpo}{B^{+0}}
\newcommand{\Boo}{B^{00}}
\newcommand{\Cpm}{C^{+-}}
\newcommand{\Spm}{S^{+-}}
\newcommand{\Coo}{C^{00}}
\newcommand{\Cpo}{C^{+0}}

\newcommand{\Tpm}{T^{+-}}
\newcommand{\Pe}{P}
\newcommand{\Too}{T^{00}}

\newcommand{\dPe}{\delta_P}
\newcommand{\dToo}{\delta_0}

\newcommand{\V}[1]{V_{#1}^{\phantom{\ast}}}
\newcommand{\Vc}[1]{V_{#1}^{\ast}}

\newcommand{\im}[1]{{\text{Im}\left[#1\right]}}
\newcommand{\re}[1]{{\text{Re}\left[#1\right]}}

\begin{document}

\begin{flushright}
IFIC/07-17\\ FTUV-07-0402
\end{flushright}

\begin{center}
\begin{Large}
{\bf Reparametrization Invariance, the controversial extraction of $\boldsymbol{\alpha}$ from $\boldsymbol{B\to\pi\pi}$ and New Physics}\\
\end{Large}
\vspace{0.5cm}
Francisco J. Botella$^{~a}$, Miguel Nebot$^{~b}$\\ \vspace{0.3cm}
{\small \emph{
$^a$ Departament de F\'{\i}sica Te\`orica and IFIC,\\ Universitat de Val\`encia-CSIC,\\ E-46100, Burjassot, Spain\\
$^b$ Centro de F\'{\i}sica Te\'orica de Part\'{\i}culas (CFTP),\\ Instituto Superior T\'ecnico,\\ P-1049-001, Lisboa, Portugal
}}
\end{center}

\begin{abstract}
The extraction of the weak phase $\alpha$ from $B\to\pi\pi$ decays has been controversial from a statistical point of view, as the frequentist vs. bayesian confrontation shows. We analyse several relevant questions which have not deserved full attention and pervade the extraction of $\alpha$. Reparametrization Invariance proves appropriate to understand those issues. We show that some Standard Model inspired parametrizations can be senseless or inadequate if they go beyond the minimal Gronau and London assumptions: the single weak phase $\alpha$ just in the $\Delta I=3/2$ amplitudes, the isospin relations and experimental data. Beside those analyses, we extract $\alpha$ through the use of several adequate parametrizations, showing that there is no relevant discrepancy between frequentist and bayesian results. The most relevant information, in terms of $\alpha$, is the exclusion of values around $\alpha\sim \pi/4$; this result is valid in the presence of arbitrary New Physics contributions to the $\Delta I=1/2$ piece.
\end{abstract}

\newpage
\section{Introduction}\label{SEC:01}

The extraction of the CP violating phase $\alpha$ \cite{Gronau:1990ka} has lead to some recent controversy confronting the results and statistical methods of two different collaborations: the frequentist approach advocated in references \cite{Charles:2006vd,Charles:2007yy} and the bayesian approach employed in reference \cite{Bona:2007qt}. In reference \cite{Charles:2006vd} J. Charles et al. presented an important criticism to the bayesian methods used by the UTfit collaboration in order to extract the angle $\alpha$ of the unitarity triangle $b$--$d$ from $\pi\pi$ and $\rho\rho$ data. The criticism relies heavily on the statistical treatment of data: frequentist vs. bayesian. The answer of the UTfit collaboration \cite{Bona:2007qt} rises some interesting points, both on the interpretation of the results and on the importance of the physical assumptions on the hadronic amplitudes. The authors of \cite{Charles:2006vd} have recently answered to this UTfit reply in \cite{Charles:2007yy}. The aim of the present work is to clarify several issues central to an adequate understanding of the physics at stake. We also want to call the attention on the importance of reparametrization invariance (RpI) in the sense introduced by F.J.B. and J. Silva in reference \cite{Botella:2005ks} to do so. We will not enter the polemic arena of statistical confrontation. With regard to this, we will instead illustrate the compatibility of results obtained in both approaches as long as things are done properly; notwithstanding, we will not ignore some ``obscure'' aspects of both approaches that are somehow swept under the rug as the statistical confrontation rages on, they illustrate that rather than sticking to one approach and deprecating the other it may be wiser to learn lessons from both.

This work is organized as follows. We start section \ref{SEC:02} with a short reminder on reparametrization invariance and its implications, then
 we use the exclusion or inclusion of $B\to\pi^0\pi^0$ data together with RpI to clarify the origin of our knowledge on $\alpha$. In section \ref{SEC:04:0} we study critically Standard Model inspired parametrizations. We devote section \ref{SEC:03} to a detailed analysis of the impact on the results of allowed ranges for some parameters. The lessons from previous sections set up the stage for an adequate extraction of $\alpha$, to which section \ref{SEC:04} is dedicated, especially in the presence of New Physics (NP) in loops. Several appendices deal with aspects left out of the main flow of the discussion. 


\section{Reparametrization invariance and $\boldsymbol{B\to\pi\pi}$}\label{SEC:02}

\subsection{Weak Phases}\label{sSEC:02:00}
We start this section with a short reminder of the findings presented in reference \cite{Botella:2005ks} concerning the parametrization of decay amplitudes and the election of weak phases. A generic parametrization of the decay amplitude of a B meson to a given final state and the CP-conjugate amplitude is the following\footnote{If the final state is $\pm 1$ CP eigenstate, $\bar A$ should include an additional $\pm 1$ factor.}:
\begin{eqnarray}
 A&=&M_1~ e^{+i\phi_1}~e^{i\delta_1}+M_2~ e^{+i\phi_2}~e^{i\delta_2}~,\notag\\
\bar A&=&M_1~ e^{-i\phi_1}~e^{i\delta_1}+M_2~ e^{-i\phi_2}~e^{i\delta_2}~,\label{EQ:0200:01}
\end{eqnarray}
where $\phi_j$ are CP-odd weak phases, $\delta_j$ are CP-even strong phases and $M_j$ the magnitudes of the different contributions. The first property to consider is the full generality, as long as $\phi_1-\phi_2\neq 0 \mod [\pi]$, of \eq{EQ:0200:01}, i.e. any additional contribution $M_3e^{\pm i\phi_3}e^{i\delta_3}$ can be recast into the previous form as
\begin{equation}
 e^{\pm i\phi_3}=\frac{\sin(\phi_3-\phi_2)}{\sin(\phi_1-\phi_2)}~e^{\pm i\phi_1}+\frac{\sin(\phi_3-\phi_1)}{\sin(\phi_2-\phi_1)}~e^{\pm i\phi_2}~,\label{EQ:0200:02}
\end{equation}
and thus
\begin{eqnarray}
 A^\prime&=&A+M_3e^{+i\phi_3}e^{i\delta_3}=M_1^\prime e^{+i\phi_1}e^{i\delta^\prime_1}+M_2^\prime e^{+i\phi_2}e^{i\delta^\prime_2}~,\notag\\
\bar A^\prime&=&\bar A+M_3e^{-i\phi_3}e^{i\delta_3}=M_1^\prime e^{-i\phi_1}e^{i\delta^\prime_1}+M_2^\prime e^{-i\phi_2}e^{i\delta^\prime_2}~,\label{EQ:0200:03}
\end{eqnarray}
with 
\begin{eqnarray}
 M_1^\prime e^{i\delta^\prime_1}&=&M_1 e^{i\delta_1}+M_3e^{i\delta_3}\frac{\sin(\phi_3-\phi_2)}{\sin(\phi_1-\phi_2)}~,\notag\\
 M_2^\prime e^{i\delta^\prime_2}&=&M_2 e^{i\delta_2}+M_3e^{i\delta_3}\frac{\sin(\phi_3-\phi_1)}{\sin(\phi_2-\phi_1)}~.\label{EQ:0200:04}
\end{eqnarray}
We can also use \eq{EQ:0200:02} to change our basic set $\{\phi_1,\phi_2\}$ of weak phases to any other arbitrary set of weak phases $\{\varphi_1,\varphi_2\}$, as long as $\varphi_1-\varphi_2\neq 0 \mod [\pi]$:
\begin{eqnarray}
 A&=&\mathcal M_1~ e^{+i\varphi_1}~e^{i\Delta_1}+\mathcal M_2~ e^{+i\varphi_2}~e^{i\Delta_2}~,\notag\\
\bar A&=&\mathcal M_1~ e^{-i\varphi_1}~e^{i\Delta_1}+\mathcal M_2~ e^{-i\varphi_2}~e^{i\Delta_2}~,\label{EQ:0200:05}
\end{eqnarray}
where
\begin{eqnarray}
 \mathcal M_1e^{i\Delta_1}&=& M_1 e^{i\delta_1}\frac{\sin(\phi_1-\varphi_2)}{\sin(\varphi_1-\varphi_2)}+M_2 e^{i\delta_2}\frac{\sin(\phi_2-\varphi_2)}{\sin(\varphi_1-\varphi_2)}~,\notag\\
 \mathcal M_2e^{i\Delta_2}&=& M_1 e^{i\delta_1}\frac{\sin(\phi_1-\varphi_1)}{\sin(\varphi_2-\varphi_1)}+M_2 e^{i\delta_2}\frac{\sin(\phi_2-\varphi_1)}{\sin(\varphi_2-\varphi_1)}~.
\end{eqnarray}
This change in the basic set of chosen weak phases should have no physical implications, hence the name \emph{reparametrization invariance}. We remind two main consequences of RpI in the absence of hadronic inputs. For an extensive discussion see \cite{Botella:2005ks}:
\begin{enumerate}
 \item Consider two basic sets of weak phases $\{\phi_1,\phi_2\}$ and $\{\phi_1,\varphi_2\}$ with $\phi_2\neq\varphi_2$; if an algorithm allows us to write $\phi_2$ as a function of physical observables then, owing to the functional similarity of equation (\ref{EQ:0200:01}) and (\ref{EQ:0200:05}), we would extract $\varphi_2$ with exactly the same function, leading to $\phi_2=\varphi_2$, in contradiction with the assumptions; then, a priori, the weak phases in the parametrization of the decay amplitudes have no physical meaning, or cannot be extracted without hadronic input.
 \item If, experimentally, the direct CP asymmetry $C=(|A|^2-|\bar A|^2)/(|A|^2+|\bar A|^2)$ is $C=0$, then the decay amplitudes can be expressed in terms of a single weak phase, which could be sensibly extracted, up to discrete ambig\"uities, through the indirect CP asymmetry $S=2~\text{Im}(\bar A A^\ast)/(|A|^2+|\bar A|^2)$. Additionally, if the theoretical description of the decay amplitudes only involves a single weak phase from a basic Lagrangian, then it can be identified with the phase measured through $S$.
\end{enumerate}
As we will see, this two results apply respectively to the $\pi^+\pi^-$ and $\pi^+\pi^0$ channels. Essentially, the first one  will be operative in the $\Delta I=1/2$ piece and the second one in the $\Delta I=3/2$.

\subsection{Removing $\boldsymbol{\pi^0\pi^0}$ information}\label{sSEC:02:01}

 To make our point transparent we will start by studying the extraction -- in fact the \emph{non-extraction} -- of $\alpha$ from $\pi\pi$ data when $B\to\pi^0\pi^0$ experimental information is removed. Let us start with a widely used \cite{Charles:2006vd,Bona:2007qt}, Standard Model inspired, parametrization of the decay amplitudes:
\begin{align}
\Apm &\equiv  A(\Bd\to\pi^+\pi^-)=e^{-i\alpha}\Tpm+\Pe~, \notag\\
\sqrt 2\Apo &\equiv  \sqrt 2 A(\Bp\to\pi^+\pi^0)=e^{-i\alpha}(\Tpm+\Too)~,\notag\\
\sqrt 2\Aoo &\equiv  \sqrt 2 A(\Bd\to\pi^0\pi^0)\equiv\sqrt 2\Apo-\Apm=e^{-i\alpha}\Too-\Pe ~,\notag\\
\ApmB &\equiv  A(\Bbd\to\pi^+\pi^-)=e^{+i\alpha}\Tpm+\Pe~, \notag\\
\sqrt 2\ApoB &\equiv  \sqrt 2 A(\Bm\to\pi^-\pi^0)=e^{+i\alpha}(\Tpm+\Too)~,\notag\\
\sqrt 2\AooB &\equiv  \sqrt 2 A(\Bbd\to\pi^0\pi^0)\equiv\sqrt 2\ApoB-\ApmB=e^{+i\alpha}\Too-\Pe ~.
\label{EQ:SMparam}
\end{align}
When $\pi^0\pi^0$ experimental information is removed we have two \emph{decoupled} decays:
\begin{enumerate}
 \item $\pi^+\pi^0$ data, i.e. the average branching ratio $\Bpo$ and the direct CP asymmetry $\Cpo$, provide, respectively, $|\Tpm+\Too|$ and a consistency check $\Cpo=0$; $\alpha$ is \emph{irrelevant} there.
 \item $\pi^+\pi^-$ data, i.e. $\Bpm$, $\Cpm$ and the mixing induced CP asymmetry $\Spm$, give information on $\alpha$ \emph{decoupled} from $\pi^+\pi^0$, on $|\Tpm|$, $|\Pe|$ and the relative (strong) phase $\delta_{\Pe\Tpm}$ between $\Tpm$ and $\Pe$.
\end{enumerate}

\begin{figure}[h]
\begin{center}
\subfigure[$\alpha$ PDF\label{FIG:alpha:alphaprime:01a}]{\epsfig{file=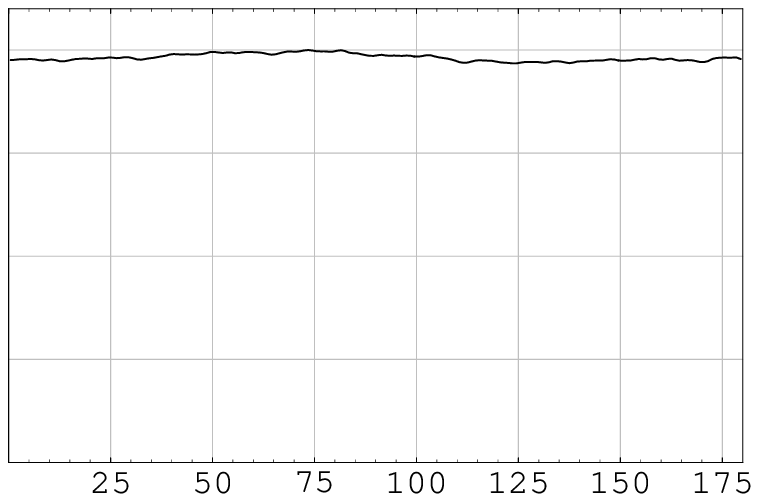,width=0.45\textwidth}}\quad
\subfigure[$\alpha^\prime$ PDF\label{FIG:alpha:alphaprime:01b}]{\epsfig{file=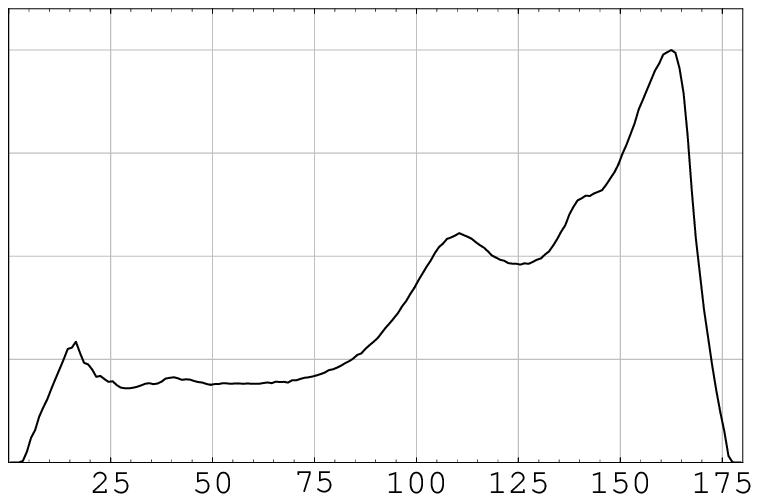,width=0.45\textwidth}}\\
\subfigure[Joint $(\alpha^\prime,\alpha)$ PDF\label{FIG:alpha:alphaprime:01c}]{\epsfig{file=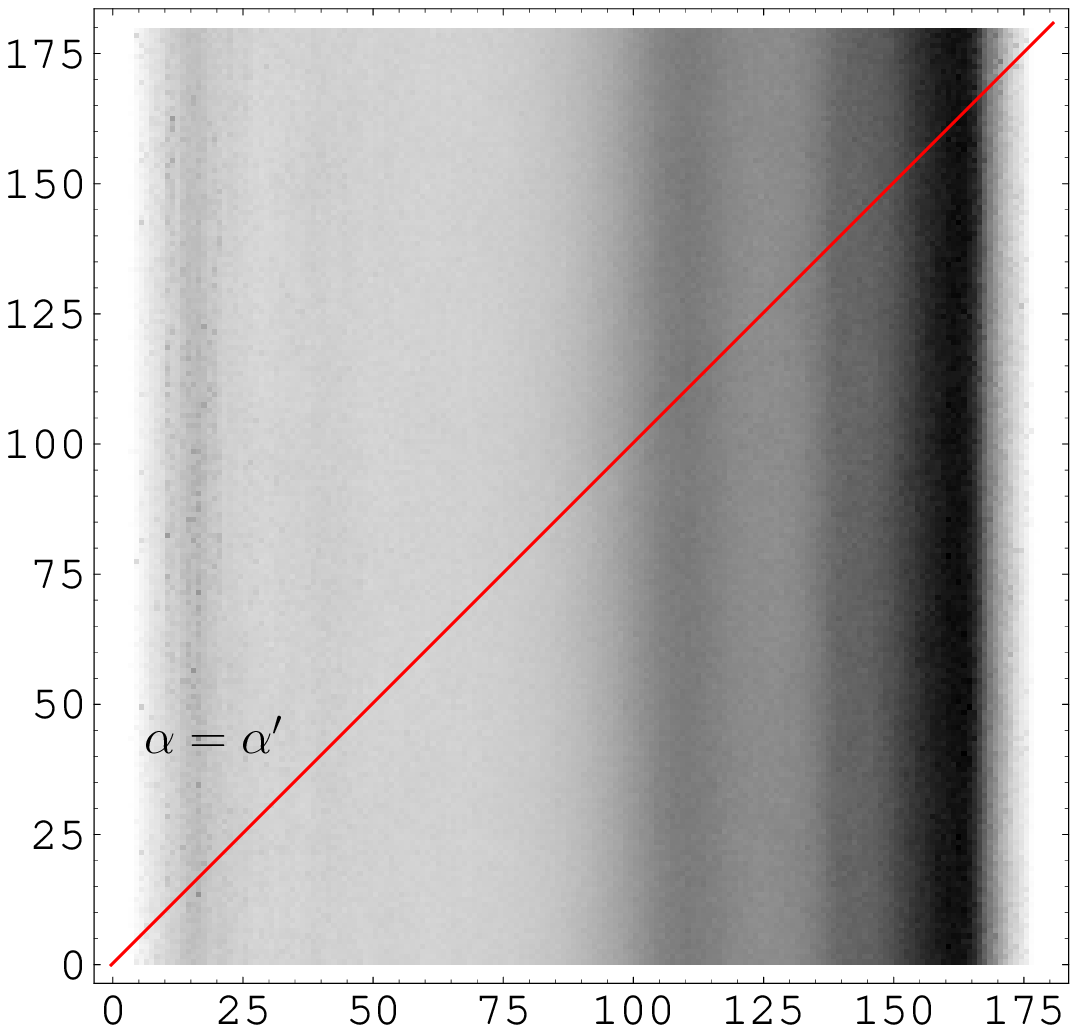,width=0.32\textwidth}}\quad
\subfigure[$\alpha=\alpha^\prime$ PDF\label{FIG:alpha:alphaprime:01d}]{\epsfig{file=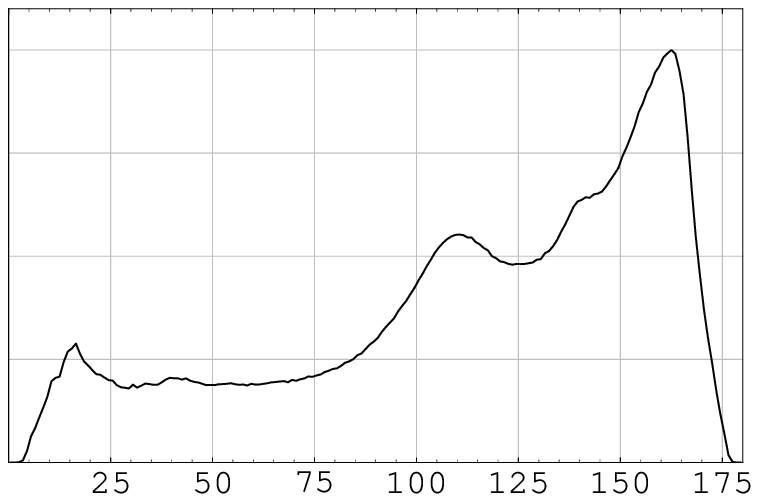,width=0.5\textwidth}}
\caption{PDFs of $\alpha$ and $\alpha^\prime$ from $B\to\pi\pi$ without $\pi^0\pi^0$ data.}\label{FIG:alpha:alphaprime:01}
\end{center}
\end{figure}

With three observables and four parameters everybody knows or suspects that one cannot really extract $\alpha$: we have $\Cpm\neq 0$, as reminded in section \ref{sSEC:02:00}, $\alpha$ \emph{cannot} be extracted from $B\to\pi^+\pi^-$ in this limited case. One can try, nevertheless, to obtain a probability distribution function (PDF) for $\alpha$ as in reference \cite{Charles:2006vd}. This PDF, obtained in an analysis with three observables and four unknowns, has obviously a strong dependence in the priors, as in figure 2 of \cite{Charles:2006vd}. Even worse, reparametrization invariance \cite{Botella:2005ks} tells us that $\Apm,\ApmB$ can also be written as
\begin{equation}
 \Apm=e^{-i\alpha^\prime}T^{\prime +-}+P^\prime,\qquad \ApmB=e^{+i\alpha^\prime}T^{\prime +-}+P^\prime~,
\end{equation}
where $\alpha^\prime$ is any weak phase -- known or unknown, $\alpha^\prime\neq 0,\pi$ --. In this scenario the conclusion is clear: \emph{any} information one would get for $\alpha$ would also be valid for \emph{any} $\alpha^\prime$ and thus it cannot be assigned to $\alpha$. This solves the puzzle raised in the MA and RI parametrizations within figure 4 of reference \cite{Charles:2006vd}: those PDFs cannot be attributable  to $\alpha$. Just with that data alone we cannot extract $\alpha^\prime$ -- whatever it is --, as we have emphasized in \ref{sSEC:02:00}. To illustrate this issue we compute the PDFs of figure \ref{FIG:alpha:alphaprime:01} in the following parametrization:
\begin{align}
\Apm &\equiv  A(\Bd\to\pi^+\pi^-)=e^{-i\alpha^\prime}\Tpm+\Pe~, \notag\\
\sqrt 2\Apo &\equiv  \sqrt 2~ A(\Bp\to\pi^+\pi^0)=e^{-i\alpha}(\Tpm+\Too)~,\notag\\
\sqrt 2\Aoo &\equiv  \sqrt 2~ A(\Bd\to\pi^0\pi^0)\equiv\sqrt 2\Apo-\Apm~, \notag\\
\ApmB &\equiv  A(\Bbd\to\pi^+\pi^-)=e^{+i\alpha^\prime}\Tpm+\Pe~, \notag\\
\sqrt 2\ApoB &\equiv  \sqrt 2~ A(\Bm\to\pi^-\pi^0)=e^{+i\alpha}(\Tpm+\Too)~,\notag\\
\sqrt 2\AooB &\equiv  \sqrt 2~ A(\Bbd\to\pi^0\pi^0)\equiv\sqrt 2\ApoB-\ApmB~. \notag\\
\label{EQ:AlphaPrime:Parametrization}
\end{align}
Notice that just with $\alpha=\alpha^\prime$, \eq{EQ:AlphaPrime:Parametrization} recovers the parametrization in \eq{EQ:SMparam}. The phase of $\Tpm$ is set to zero (i.e. all strong phases are relative to $\arg(\Tpm)$) and flat priors are used for all the parameters\footnote{The allowed ranges for the different moduli and the sensitivity to them in this and other cases will be addressed later, for instance, for this example, they are all limited to lie in the range $[0;10]\times 10^{-3}\text{ ps}^{-1/2}$.}, that is, moduli $|\Tpm|$, $|\Too|$, $|\Pe|$ and phases $\dPe=\arg(\Pe)$, $\dToo=\arg(\Too)$, $\alpha$ and $\alpha^\prime$. Results in other parametrizations, being equally illustrative, are relegated to appendix \ref{AP:03}.

The lesson of this example is rather obvious: the set of observables being insensitive to $\alpha$, its PDF is uninformative (just the flat prior in this case); the PDF in figure \ref{FIG:alpha:alphaprime:01d}, erroneously identified with $\alpha$, is nothing else than $\alpha^\prime$ itself, whatever it could be.


\subsection{Including back $\boldsymbol{\pi^0\pi^0}$ information}\label{sSEC:02:02}
\begin{figure}[h]
\begin{center}
\subfigure[$\alpha$ PDF\label{FIG:alpha:alphaprime:02a}]{\epsfig{file=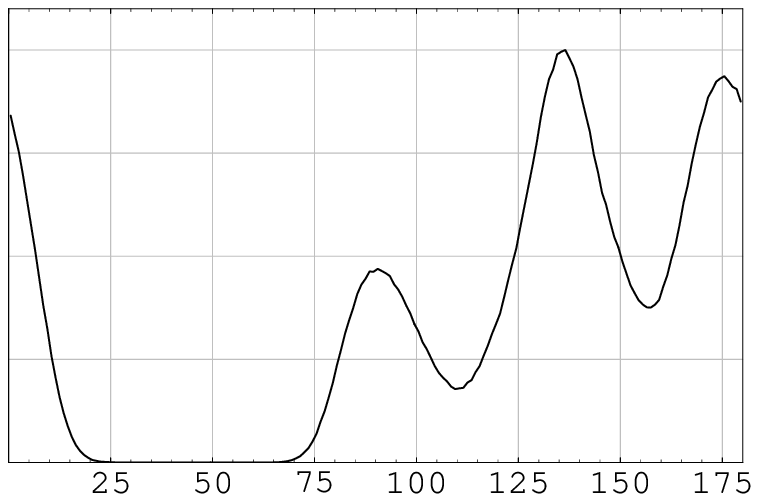,width=0.45\textwidth}}\quad
\subfigure[$\alpha^\prime$ PDF\label{FIG:alpha:alphaprime:02b}]{\epsfig{file=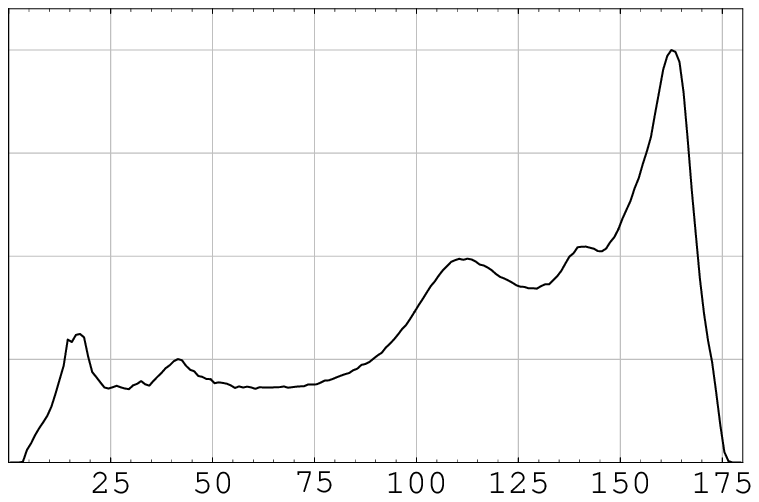,width=0.45\textwidth}}\\
\subfigure[Joint $(\alpha^\prime,\alpha)$ PDF\label{FIG:alpha:alphaprime:02c}]{\epsfig{file=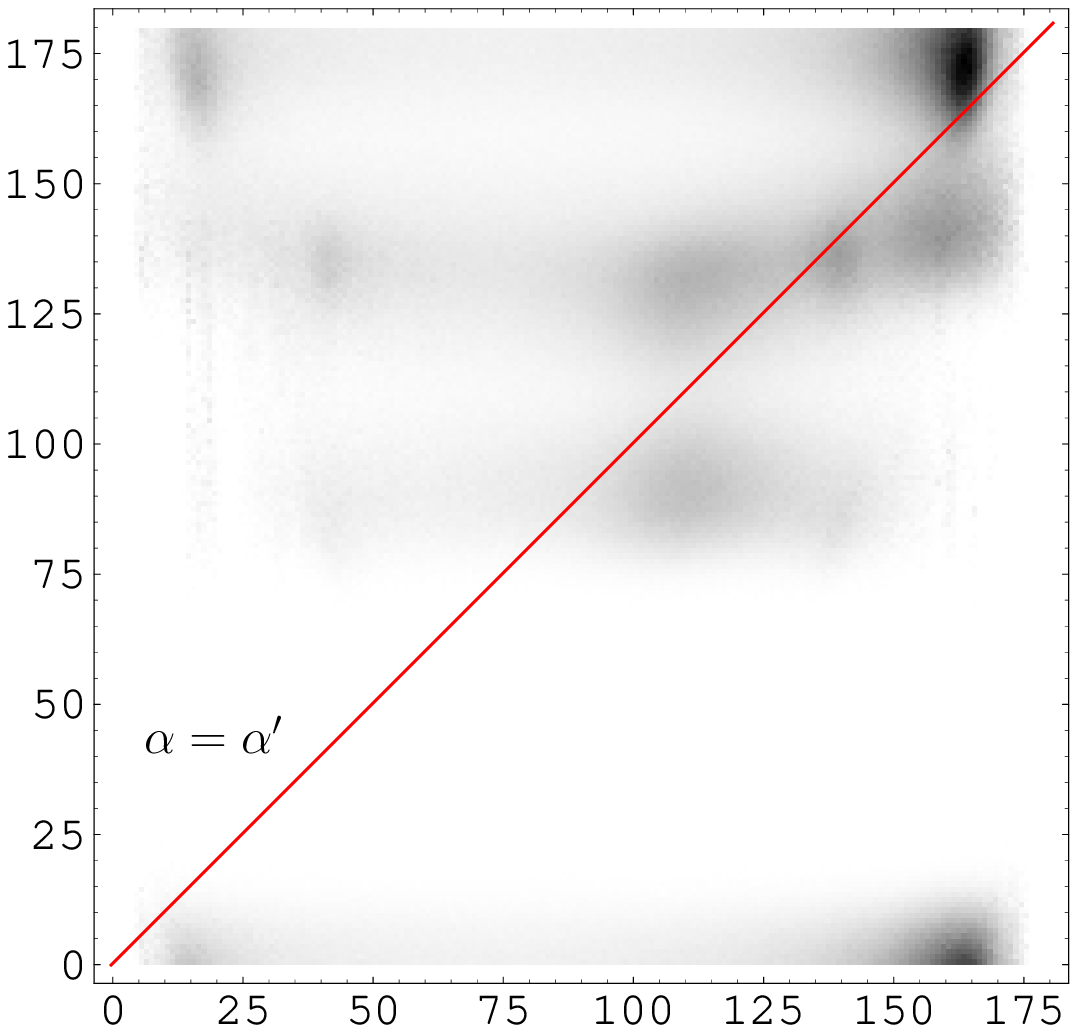,width=0.32\textwidth}}\quad
\subfigure[$\alpha=\alpha^\prime$ PDF\label{FIG:alpha:alphaprime:02d}]{\epsfig{file=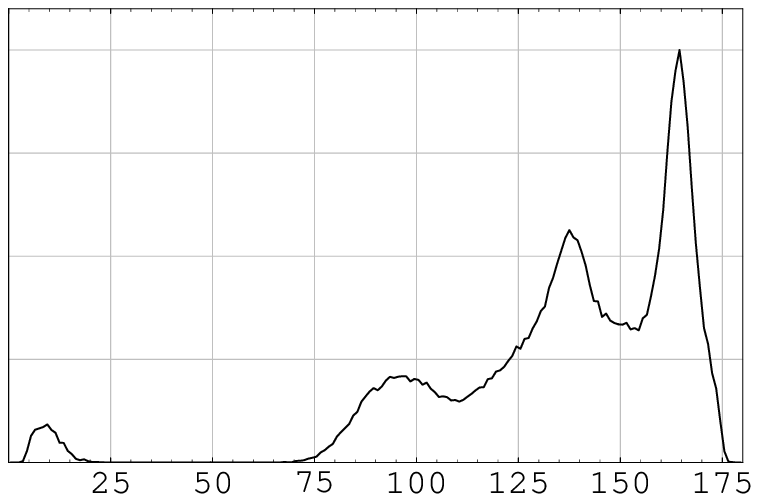,width=0.5\textwidth}}
\caption{PDFs of $\alpha$ and $\alpha^\prime$ from $B\to\pi\pi$.}\label{FIG:alpha:alphaprime:02}
\end{center}
\end{figure}

When we incorporate $B\to\pi^0\pi^0$ data to the isospin construction, $|\Aoo|$ ($|\AooB|$) gives the angle among $\Apo$ ($\ApoB$) and $\Apm$ ($\ApmB$); using then the known phase difference between $\Apm$ and $\ApmB$, the angle among $\Apo$ and $\ApoB$ is obtained. This is just the isospin analysis giving $\alpha$. Knowing $\alpha$, i.e. with $\alpha$ fixed, $\Apm=e^{-i\alpha}\Tpm+\Pe$ would have full meaning and $\{B_{+-},C_{+-},S_{+-}\}$ would fix the three hadronic parameters. Unfortunately the isospin analysis as explained above yields allowed values for $\alpha$ spanning a wide range. The degeneracy of solutions together with the experimental errors do not fix $\alpha$, just exclude some region. In this situation $\{B_{+-},C_{+-},S_{+-}\}$ do not really fix the hadronic parameters and, consequently, they tend to generate a spurious PDF for $\alpha$ as we have seen. The final ``$\alpha$'' is thus a sort of convolution of the $\alpha$ obtained from the isospin analysis and the spurious one ``extracted'' purely from $\pi^+\pi^-$ data.
This is illustrated with the PDFs of figure \ref{FIG:alpha:alphaprime:02}, making use of the parametrization in \eq{EQ:AlphaPrime:Parametrization}.  

\begin{figure}[h]
\begin{center}
\subfigure[$\alpha$ PDF]{\epsfig{file=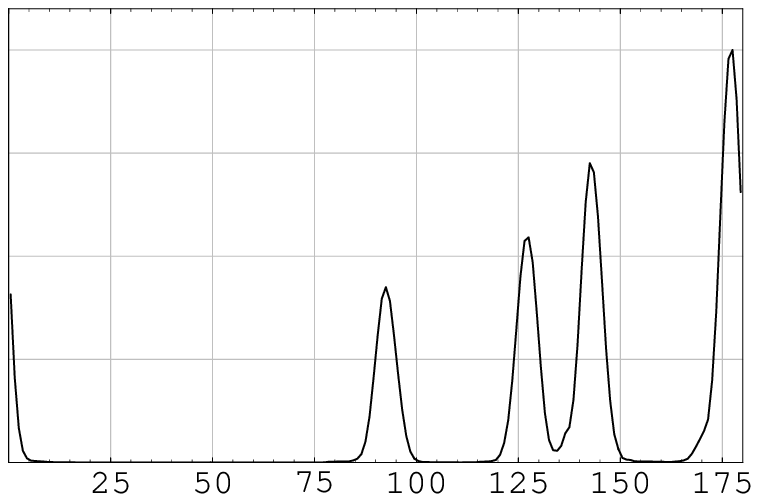,width=0.3\textwidth}}\quad
\subfigure[$\alpha^\prime$ PDF]{\epsfig{file=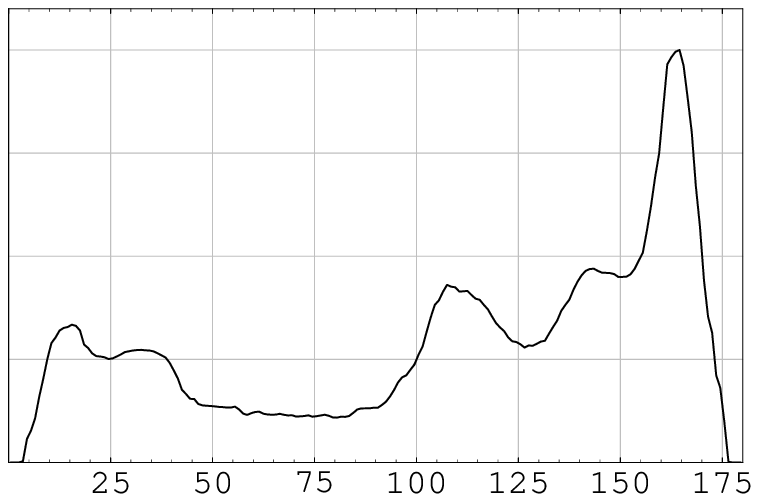,width=0.3\textwidth}}\quad
\subfigure[$\alpha=\alpha^\prime$ PDF]{\epsfig{file=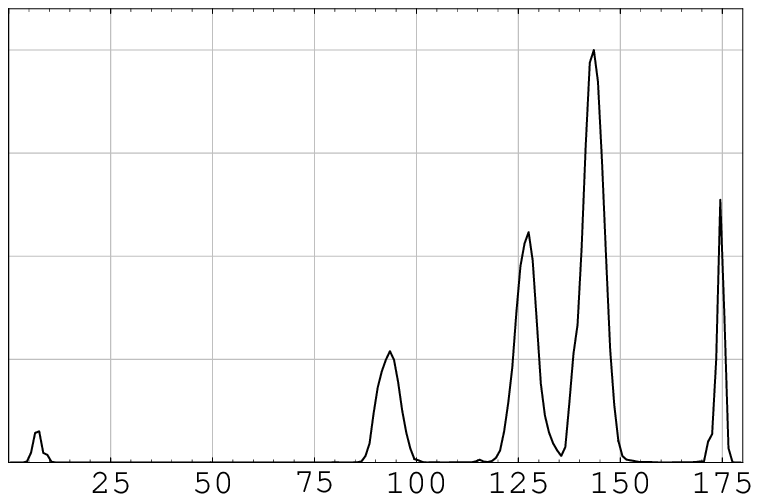,width=0.3\textwidth}}
\caption{PDFs of $\alpha$ and $\alpha^\prime$ from $B\to\pi\pi$ with experimental uncertainties reduced by a factor of 5.} \label{FIG:alpha:alphaprime:03}
\end{center}
\end{figure}

To stress the importance of this issue we repeat the previous example while arbitrarily reducing \emph{all} experimental uncertainties by a common factor of 5. The PDFs corresponding to this fake scenario are displayed in figure \ref{FIG:alpha:alphaprime:03}.

The results shown in figures \ref{FIG:alpha:alphaprime:01}, \ref{FIG:alpha:alphaprime:02} and \ref{FIG:alpha:alphaprime:03} deserve some comment:
\begin{enumerate}
 \item Figures \ref{FIG:alpha:alphaprime:01b} and \ref{FIG:alpha:alphaprime:02b} are almost identical; in the former we were not using $B\to\pi^0\pi^0$ information while in the later we were doing so. This similarity is a dramatic illustration of the spurious nature of the ``extracted'' $\alpha^\prime$.
 \item Figure \ref{FIG:alpha:alphaprime:02d} is the cut of the joint PDF in figure \ref{FIG:alpha:alphaprime:02c} along the line $\alpha=\alpha^\prime$. Therefore the so called MA extraction of $\alpha$ is a sort of convolution of the Gronau-London $\alpha$ -- figure \ref{FIG:alpha:alphaprime:02a} -- and the spurious one.
 \item This $\alpha^\prime$ PDF basically allows any value of $\alpha^\prime$ except the neighborhoods of $0$ and $\pi$, which are a priori forbidden by $\Spm,\Cpm\neq 0$: obviously there is no way to produce CP violation in the $\pi^+\pi^-$ channel without two weak phases in the amplitude that controls it. The exclusion of $\alpha^\prime=0,\pi$ is the only physical information one can extract in the SM from the PDF of $\alpha^\prime$.
 \item The deep in the $\alpha$ distributions around $\alpha\sim \pi/4$, which is transmitted to the $\alpha=\alpha^\prime$ PDF, is senseful. The exclusion of $\alpha\sim 0,\pi$ is also physical inside the SM. Nevertheless, how strongly these $0,\pi$ regions are excluded is highly sensitive to the allowed ranges for $|\Tpm|$, $|\Too|$ and $|\Pe|$ -- see section \ref{SEC:03} --. As we move away form the $\alpha=0,\pi$ points, the final PDF of $\alpha$ would be more influenced by the spurious $\alpha^\prime$ distribution. One can see that in the shape of the $\alpha$ distribution for $\alpha<25^\circ$ or $\alpha>75^\circ$.
 \item As uncertainties are reduced, even with $\alpha\equiv\alpha^\prime$, the valid ranges for the ``real'' $\alpha$ emerge, despite the $\alpha^\prime$ distribution. That is, as experimental uncertainties are reduced, the $\alpha^\prime$ ``pollution'' of $\alpha$ through $\alpha\equiv\alpha^\prime$ becomes increasingly ineffective, as it should, and just transmits the physical exclusion of $\alpha=0,\pi$ inside the SM.
\end{enumerate}

The main lesson from the previous example is: $\alpha$ is obtained from purely $\Delta I=3/2$ amplitudes, without additional \emph{hadronic} input. Including it in $\Delta I=1/2$ pieces, as reparametrization invariance shows, pollutes the legitimate extraction with information that one cannot claim is concerning $\alpha$.

\section{Standard Model inspired parametrizations}\label{AP:05}\label{SEC:04:0}
 As stated above, following the consequences of reparametrization invariance, the really legitimate sources of our knowledge on $\alpha$ are $\Apo,\ApoB$. We have referred to the parametrization in \eq{EQ:SMparam} as a ``SM inspired parametrization'' of the amplitudes and we have discussed how the inclusion of $\alpha$ in $\Apm,\ApmB$ is dangerous with present uncertainties. Nevertheless, it is clear that the exclusion of $\alpha\sim 0,\pi$ inside the SM is a valid physical consequence that comes from having $\alpha$ in $\Apm$ and $\ApmB$. To further illustrate the importance and the subtlety of this issue let us consider in detail what can be interpreted as a ``SM inspired parametrization''. Once we take into account reparametrization invariance, we only need\footnote{$\Apo$ and $\ApoB$ can be parametrized with a single weak phase, identifiable with $\alpha$, $\Aoo$ and $\AooB$ will follow from the isospin relations.} to focus on $\Apm$ and $\ApmB$:
\begin{enumerate}
 \item RpI allows us to write $\{\Apm,\ApmB\}$ in terms of any pair of weak phases $\{\phi_1,\phi_2\}$ (as long as $\phi_1-\phi_2\neq 0\mod[\pi]$), \emph{nothing} enforces the use of $\{0,\alpha\}$.
 \item SM compliance of any parametrization only requires that the vanishing of all the SM phases leads to no CP violation, once again \emph{nothing} singles out or requires the use of $\{0,\alpha\}$.
\end{enumerate}
Consequently, as we have at our disposal other SM phases that we can choose to parametrize $\Apm,\ApmB$, namely\footnote{$\gamma=\arg(-\V{ud}\V{cb}\Vc{ub}\Vc{cd})$, $\beta=\arg(-\V{cd}\V{tb}\Vc{cb}\Vc{td})$, $\chi=\arg(-\V{cb}\V{ts}\Vc{cs}\Vc{tb})$ and $\chi^\prime=\arg(-\V{us}\V{cd}\Vc{ud}\Vc{cs})$ \cite{Botella:2002fr}.} $\gamma$, $\beta$, $\chi$, $\chi^\prime$, instead of $\Apm=e^{-i\alpha}\Tpm+\Pe$ and $\ApmB=e^{i\alpha}\Tpm+\Pe$, we can for example write, on equal footing,
\begin{equation}
 \Apm=M_1 e^{i\delta_1}e^{-i\chi}+M_2e^{i\delta_2} e^{-i\beta},\qquad \ApmB=M_1 e^{i\delta_1}e^{+i\chi}+M_2e^{i\delta_2} e^{+i\beta}~,
\end{equation}
or
\begin{equation}
 \Apm=e^{-i\chi}\Tpm+\Pe,\qquad \ApmB=e^{+i\chi}\Tpm+\Pe~.\label{EQ:Apm:chi1}
\end{equation}
Within the SM $\chi\sim\mathcal O(\lambda^2)$, had we used this last parametrization (\eq{EQ:Apm:chi1}), we would have found extreme compatibility problems\footnote{Just look, for example, to the $\mathcal O(\lambda^2)\sim 2-3^\circ$ region of the different $\alpha^\prime$ PDFs in the plots of previous sections \cite{Aguilar-Saavedra:2004mt}.} that would be absent with another SM inspired parametrization: this is a dramatic illustration of the consequences of RpI mentioned in section \ref{sSEC:02:00}.
In other words, pretending that one obtains information on SM ``theoretical'' phases just by parametrizing $\Apm$ and $\ApmB$ with them is in general senseless. In this case we would have obtained that figure \ref{FIG:alpha:alphaprime:02b} is the PDF of the phase $\chi$, the one that appears in $B_s$--$\bar B_s$ mixing \cite{Ligeti:2006pm,Ball:2006xx,Grossman:2006ce,Bona:2006sa,Charles:2006yw,Botella:2006va}.

\section{Physics and parametrical problems}\label{SEC:03}

\begin{figure}[h]
\begin{center}
\subfigure[$\alpha$ PDF\label{FIG:PDFscaling:01:a}]{\epsfig{file=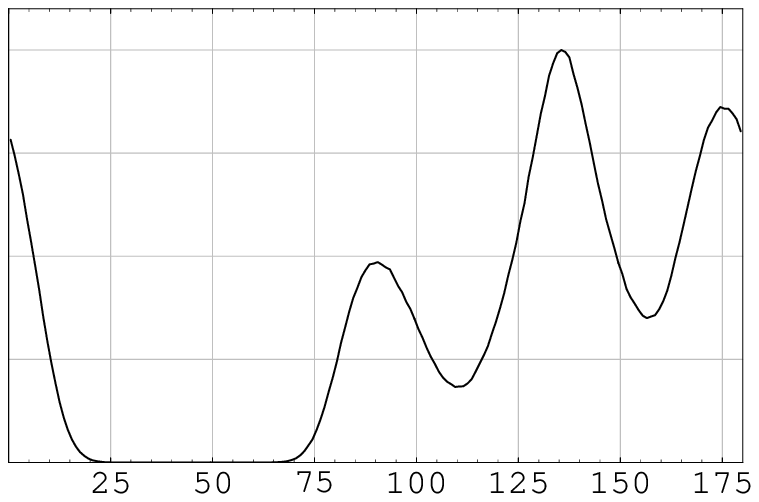,width=0.25\textwidth}}\quad
\subfigure[$\alpha^\prime$ PDF\label{FIG:PDFscaling:01:b}]{\epsfig{file=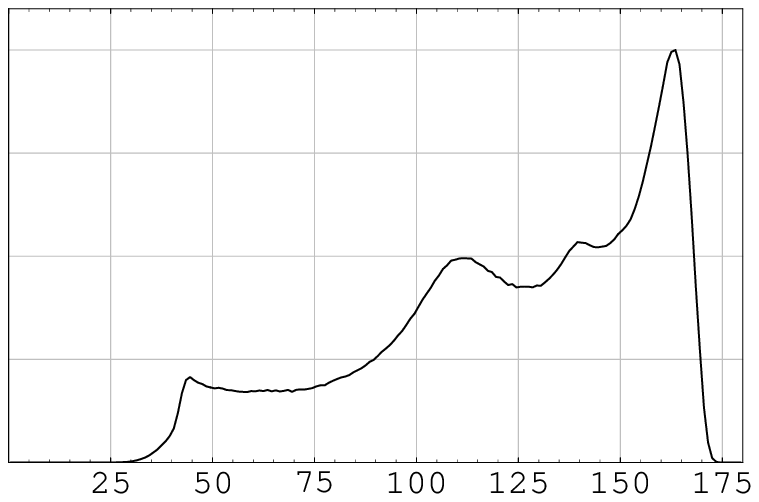,width=0.25\textwidth}}\quad
\subfigure[$\alpha=\alpha^\prime$ PDF\label{FIG:PDFscaling:01:c}]{\epsfig{file=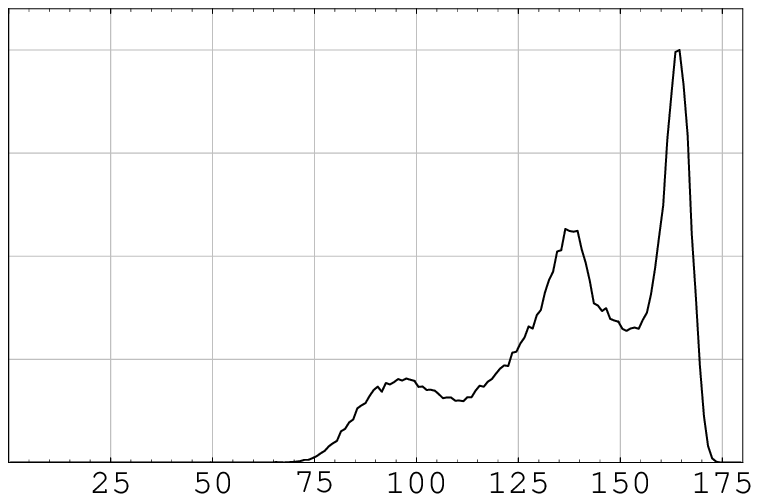,width=0.25\textwidth}}\\
Allowed ranges: $|T^{ij}|\in[0;10]\times 10^{-3}\text{ ps}^{-1/2}$, $|\Pe|\in[0;2.5]\times 10^{-3}\text{ ps}^{-1/2}$\\
\vspace{-1.2ex}\rule{0.7\textwidth}{0.5pt}\\
\subfigure[$\alpha$ PDF\label{FIG:PDFscaling:01:d}]{\epsfig{file=./Figs/fig02a.eps,width=0.25\textwidth}}\quad
\subfigure[$\alpha^\prime$ PDF\label{FIG:PDFscaling:01:e}]{\epsfig{file=./Figs/fig02b.eps,width=0.25\textwidth}}\quad
\subfigure[$\alpha=\alpha^\prime$ PDF\label{FIG:PDFscaling:01:f}]{\epsfig{file=./Figs/fig02d.eps,width=0.25\textwidth}}\\
Allowed ranges: $|T^{ij}|\in[0;10]\times 10^{-3}\text{ ps}^{-1/2}$, $|\Pe|\in[0;10]\times 10^{-3}\text{ ps}^{-1/2}$\\
\vspace{-1.2ex}\rule{0.7\textwidth}{0.5pt}\\
\subfigure[$\alpha$ PDF\label{FIG:PDFscaling:01:g}]{\epsfig{file=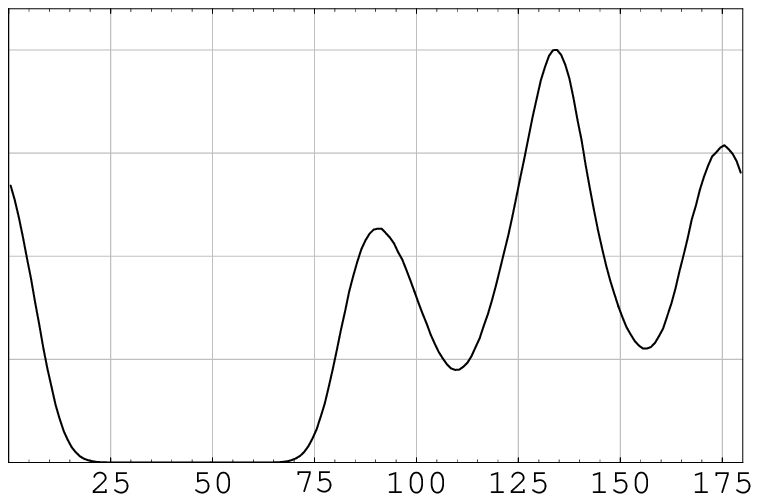,width=0.25\textwidth}}\quad
\subfigure[$\alpha^\prime$ PDF\label{FIG:PDFscaling:01:h}]{\epsfig{file=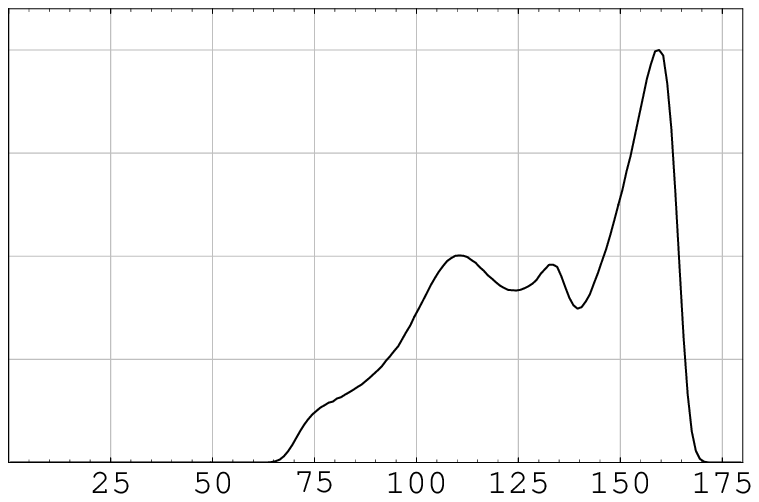,width=0.25\textwidth}}\quad
\subfigure[$\alpha=\alpha^\prime$ PDF\label{FIG:PDFscaling:01:i}]{\epsfig{file=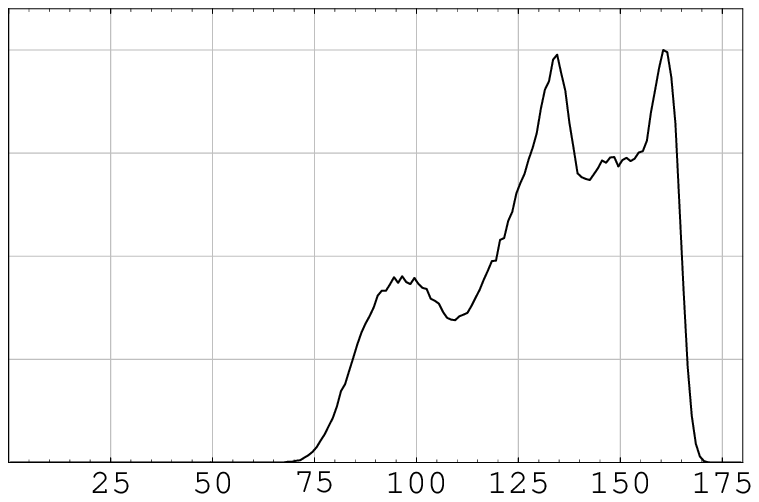,width=0.25\textwidth}}\\
Allowed ranges: $|T^{ij}|\in[0;5]\times 10^{-3}\text{ ps}^{-1/2}$, $|\Pe|\in[0;1.25]\times 10^{-3}\text{ ps}^{-1/2}$\\
\vspace{-1.2ex}\rule{0.7\textwidth}{0.5pt}\\
\subfigure[$\alpha$ PDF\label{FIG:PDFscaling:01:j}]{\epsfig{file=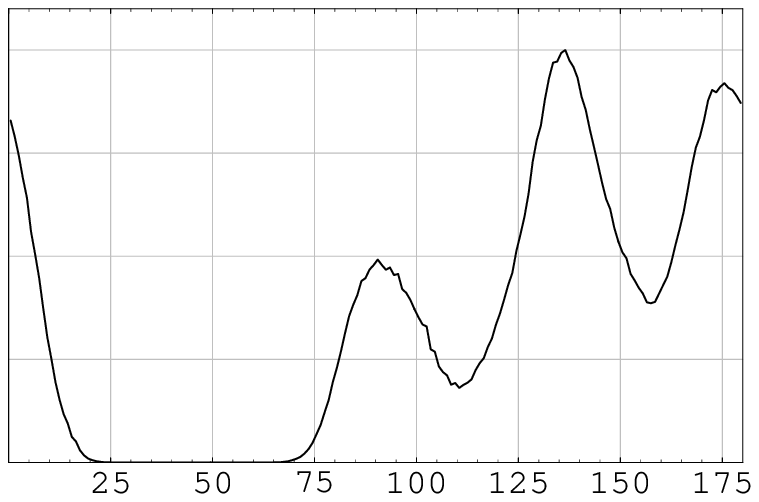,width=0.25\textwidth}}\quad
\subfigure[$\alpha^\prime$ PDF\label{FIG:PDFscaling:01:k}]{\epsfig{file=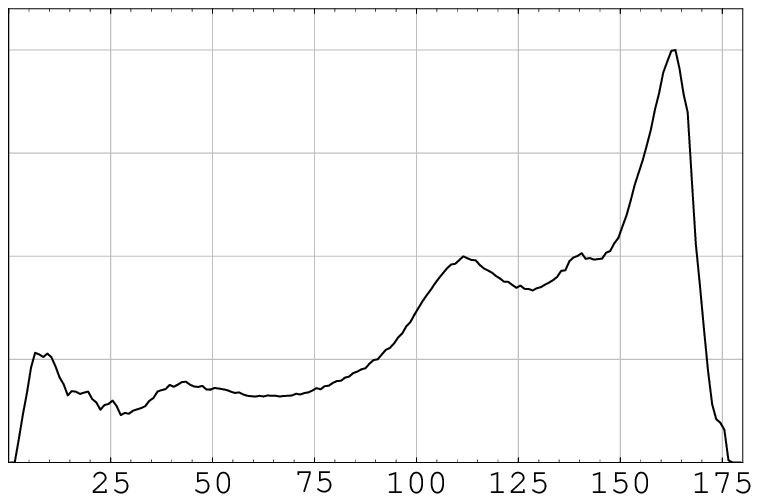,width=0.25\textwidth}}\quad
\subfigure[$\alpha=\alpha^\prime$ PDF\label{FIG:PDFscaling:01:l}]{\epsfig{file=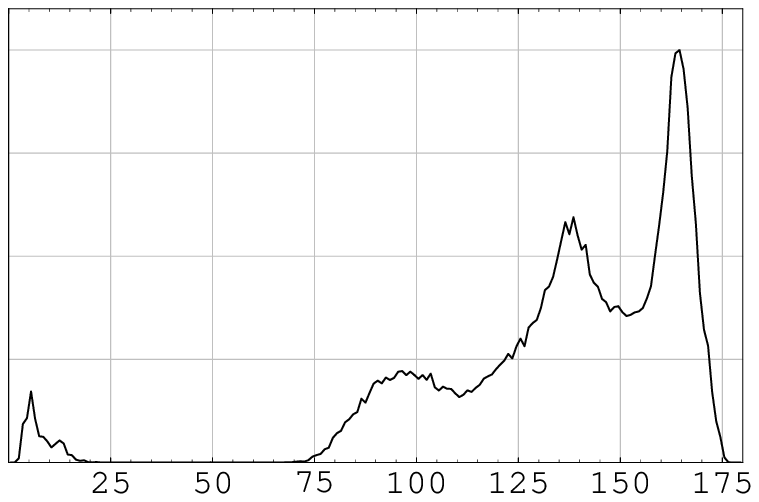,width=0.25\textwidth}}\\
Allowed ranges: $|T^{ij}|\in[0;25]\times 10^{-3}\text{ ps}^{-1/2}$, $|\Pe|\in[0;25]\times 10^{-3}\text{ ps}^{-1/2}$\\
\caption{PDFs obtained using the parametrization in \eq{EQ:AlphaPrime:Parametrization} and different allowed ranges for $|T^{ij}|$ and $|\Pe|$.} \label{FIG:PDFscaling:01}
\end{center}
\end{figure}

\begin{figure}[h]
\begin{center}
\subfigure[$|T^{ij}|<10$, $|\Pe|<2.5$\label{FIG:ScalingCL:a}]{\epsfig{file=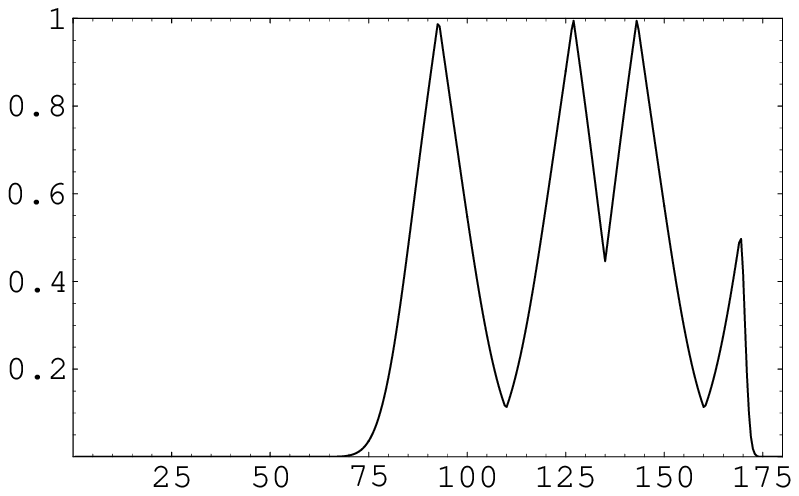,width=0.4\textwidth}}\quad
\subfigure[$|T^{ij}|<10$, $|\Pe|<10$\label{FIG:ScalingCL:b}]{\epsfig{file=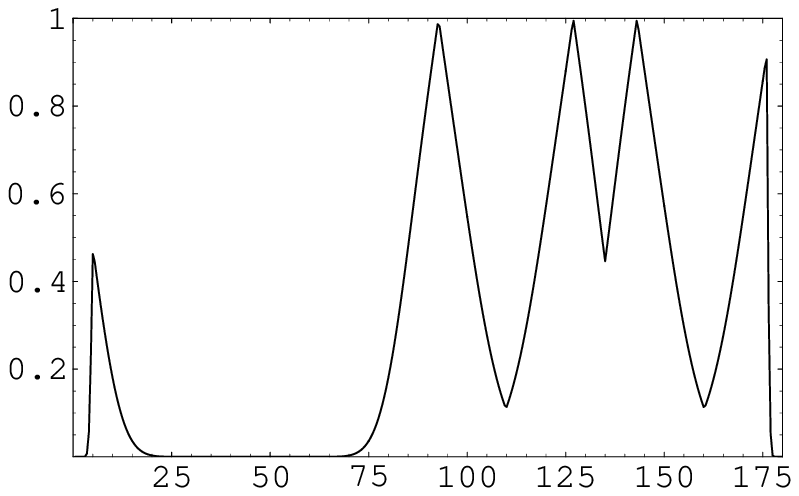,width=0.4\textwidth}}\\
\subfigure[$|T^{ij}|<5$, $|\Pe|<1.25$\label{FIG:ScalingCL:c}]{\epsfig{file=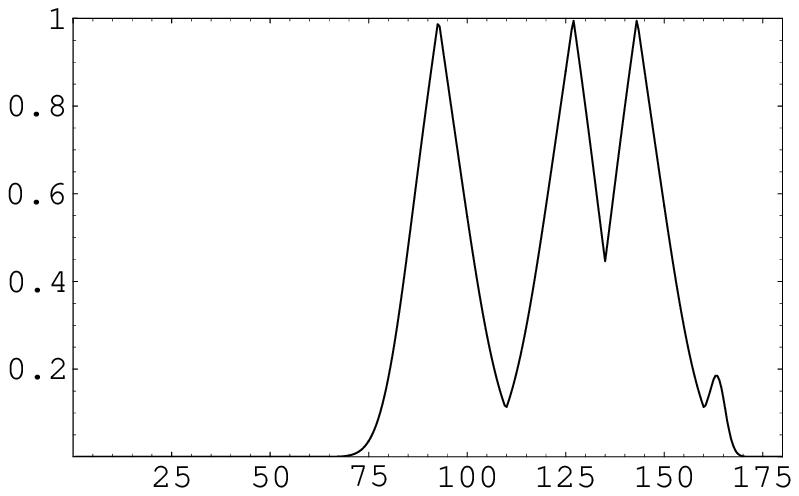,width=0.4\textwidth}}\quad
\subfigure[$|T^{ij}|<25$, $|\Pe|<25$\label{FIG:ScalingCL:d}]{\epsfig{file=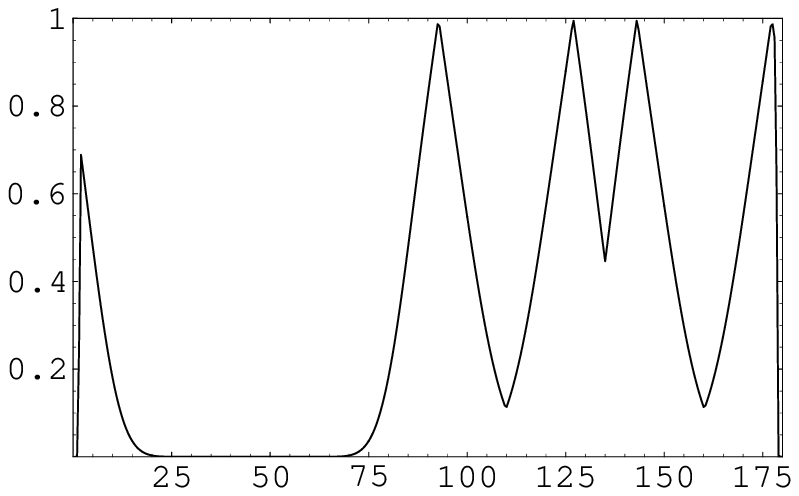,width=0.4\textwidth}}\\
\caption{$\alpha$ CL; as usual $|T^{ij}|$ and $|\Pe|$ in units of $10^{-3}\text{ ps}^{-1/2}$.} \label{FIG:ScalingCL}
\end{center}
\end{figure}

 In section \ref{SEC:02} we mentioned that the exclusion of the ``dangerous'' $\alpha^\prime$ near $0$ and $\pi$ depended on the allowed ranges for the parameters $|T^{ij}|$ and $|\Pe|$. Figure \ref{FIG:PDFscaling:01} shows the PDFs of $\alpha$, $\alpha^\prime$ and $\alpha=\alpha^\prime$ for four different sets of allowed ranges of $|T^{ij}|$ and $|\Pe|$. On the one hand, the PDFs of $\alpha$ in figures \ref{FIG:PDFscaling:01:a}, \ref{FIG:PDFscaling:01:d}, \ref{FIG:PDFscaling:01:g} and \ref{FIG:PDFscaling:01:j} are quite similar. On the other hand, the PDFs of $\alpha^\prime$ in figures \ref{FIG:PDFscaling:01:b}, \ref{FIG:PDFscaling:01:e}, \ref{FIG:PDFscaling:01:h} and \ref{FIG:PDFscaling:01:k} are completely different: the ``dangerous'' $\alpha^\prime$, especially in the regions close to $0$,$\pi$, is sensitive to the applied bounds. This is automatically transmitted to the $\alpha=\alpha^\prime$ PDF and it is in this way that the region with ``$\alpha$'' close to $0$,$\pi$ is suppressed (even wipped out as in figures \ref{FIG:PDFscaling:01:c} and \ref{FIG:PDFscaling:01:i}) through the cuts on the spurious $\alpha^\prime$, induced by the cuts on $|T^{ij}|$ and $|\Pe|$. One could think that this is particular to the bayesian statistical approach, figure \ref{FIG:ScalingCL} shows the frequentist confidence level curves for $\alpha$ computed under the same parametric restrictions. As we use the parametrization of \eq{EQ:SMparam}, they correspond to the $\alpha=\alpha^\prime$ plots in \fig{FIG:PDFscaling:01}. It is rather clear that without regard to the statistical approach, limiting the values of $|T^{ij}|$ and $|\Pe|$ has observable effects in the extraction of $\alpha$. Note that figure \ref{FIG:ScalingCL:c} differs from figure \ref{FIG:ScalingCL:a} not by a cut but by a change in the shape, even if it is not a dramatic change.

The authors of reference \cite{Charles:2006vd} pointed out that there is some peculiar limit with $\alpha\to 0$ together with $\Pe/\Tpm,\Too/\Tpm\to -1$, $|\Tpm|\to\infty$ -- using the parametrization of \eq{EQ:SMparam} -- that keeps all the observables ``in place'': it is in fact a question of having $\alpha^\prime\to 0$ rather than $\alpha\to 0$. This peculiar limit is useful to understand the $\alpha\sim 0,\pi$ exclusion above mentioned. To obtain parameter configurations with high likelihood when $\alpha^{(\prime)}$ approaches $0$ or $\pi$, the required values of $|T^{ij}|$ and $|\Pe|$ are increasingly large. Imposing bounds on $|T^{ij}|$ and $|\Pe|$ automatically limits how close to $0,\pi$ one can push the weak phase while producing likely branching ratios and asymmetries. The use of the parametrization in \eq{EQ:AlphaPrime:Parametrization} shows how this works for the dangerous $\alpha^\prime$ and is then transmitted to $\alpha$.

\clearpage
\newpage

\section{The extraction of $\boldsymbol{\alpha}$ from $\boldsymbol{B\to\pi\pi}$ and New Physics}\label{SEC:04}

Recently the UTfit collaboration has proposed to add information on the moduli of the amplitudes in order to extract $\alpha$ inside the SM. In particular, to add reasonable QCD based cuts on the moduli of $T^{ij}$ and $\Pe$. Even if we agree with this procedure, we must stress that the resulting PDF of $\alpha$ -- see figures \ref{FIG:PDFscaling:01:c} or \ref{FIG:PDFscaling:01:i} -- in the non zero region mixes $\Delta I=3/2$ information with spurious $\Delta I=1/2$ information. In this case it does not seem dramatic, but it can be so in the $B\to\rho\rho$ case -- see \cite{Charles:2006vd} --. In addition, if one is trying to make a general fit of the SM it is more natural to use the $\Delta I=3/2$ piece of $B\to\pi\pi$ to get reliable bounds on $\alpha$ and once $\alpha$ is fixed by the general unitarity triangle analysis, use the $\Delta I=1/2$ piece of $B\to\pi\pi$ to obtain better information on the hadronic parameters. In fact, the UTfit collaboration presents results along this line in \cite{Bona:2007qt}. This implies our recommendation of using $\alpha$ in the $\Apo$ amplitude and another phase in $\Apm$ or in the $\Delta I=1/2$ piece. 

After confronting the SM \emph{\`a la CKM} with data, the most important objective in overconstraining the unitarity triangle is in fact to look for New Physics (NP) \cite{Charles:2004jd,Bona:2005vz,Botella:2005fc,Ligeti:2006pm,Ball:2006xx,Grossman:2006ce,Bona:2006sa,Charles:2006yw,Botella:2006va}. When there is NP -- just in the mixings or also in the $\Delta I=1/2$ decay amplitudes\footnote{With great accuracy -- up to small electroweak penguins -- this case corresponds to having NP everywhere except in tree level amplitudes.} -- it is not appropriate to use a SM inspired parametrization. In the limit where all SM phases go to zero, $\Cpm$ and $\Spm$ can still be reproduced by NP loops. So, if we want to interpret the $\alpha$ PDF as\footnote{Where $\bar\alpha=\pi-\bar\beta-\gamma$, $\bar\beta=\beta-\phi_d$ and the NP phase in $B^0_d$--$\bar B^0_d$ mixing is defined by $M_{12}^d=r_d^2 e^{-i2\phi_d}[M_{12}^d]_{SM}$.} $\bar\alpha$ we have to use a different CP-violating phase in the $\Delta I=1/2$ piece or in $\Apm$. Parametrizations that fulfill these requirements are the so-called PLD, ES, the '$\tau$' parametrization in \cite{Charles:2006vd} and even our SM-like parametrization with $\alpha^\prime$ in \eq{EQ:AlphaPrime:Parametrization} despite having one more parameter. A similar one, which additionally factorizes an overall scale of the amplitudes, is the following, that we call '1i': 
\begin{equation}
\begin{array}{rclcrcl}
\Apm &\equiv& e^{-i\alpha}T_{3/2}(T+iP), & &\sqrt 2\Aoo &\equiv & e^{-i\alpha}T_{3/2}(1-T-iP),\\
\sqrt 2\Apo &\equiv & e^{-i\alpha}T_{3/2},& & \sqrt 2\ApoB &\equiv & e^{+i\alpha}T_{3/2},\\
\ApmB &\equiv &e^{+i\alpha}T_{3/2}(T-iP), & & \sqrt 2\AooB &\equiv  &e^{+i\alpha}T_{3/2}(1-T+iP).
\end{array}
\label{EQ:1iparam:01}
\end{equation}
Notice that a global weak phase in $\Apm$ is irrelevant in $\Cpm$ and amounts to a global shift of $\arg(\ApmB\Apm^\ast)$.

In this section we will ``extract'' $\alpha$ in a bayesian approach making use of different parametrizations; we will show the consistency of all those results and then compare to frequentist results. From a fundamental point of view, as stressed in previous sections, we are not willing to use information beside assuming the triangular isospin relations, the single ``tree level'' weak phase of the $\Delta I=3/2$ piece and experimental results themselves. Reparametrization invariance and the presence of a single weak phase, $\alpha$, in the $\Delta I=3/2$ amplitudes $\Apo$ and $\ApoB$ imply that all the results to be presented in this section will be valid in the presence of New Physics in loops.

\begin{figure}[h]
\begin{center}
\subfigure[PLD parametrization\label{FIG:Alpha:01:a}]{\epsfig{file=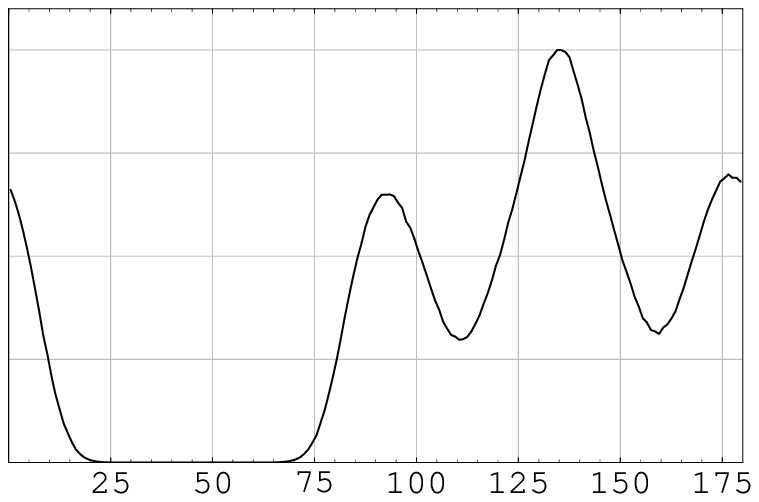,width=0.45\textwidth}}\quad
\subfigure[1i parametrization\label{FIG:Alpha:01:b}]{\epsfig{file=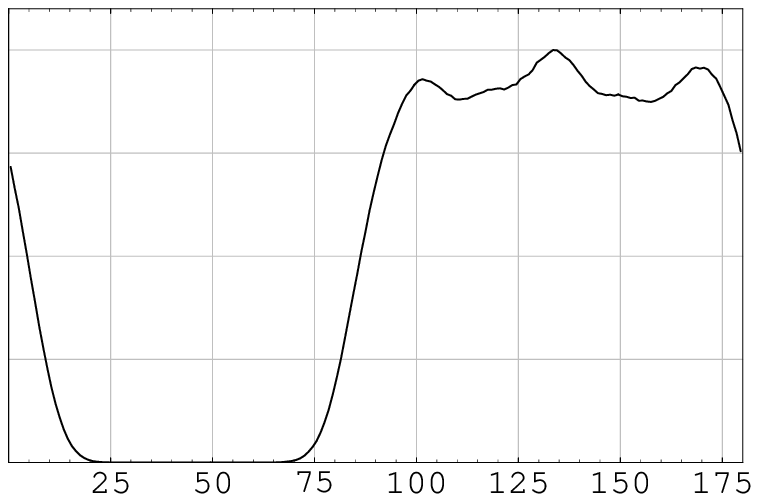,width=0.45\textwidth}}\\
\subfigure[Explicit extraction\label{FIG:Alpha:01:c}]{\epsfig{file=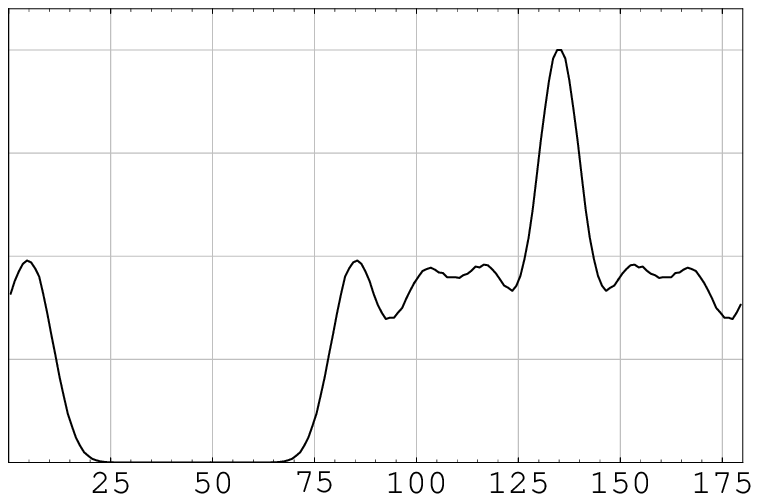,width=0.45\textwidth}}
\caption{$\alpha$ PDFs.} \label{FIG:Alpha:01}
\end{center}
\end{figure}

\begin{figure}[h]
\begin{center}
\epsfig{file=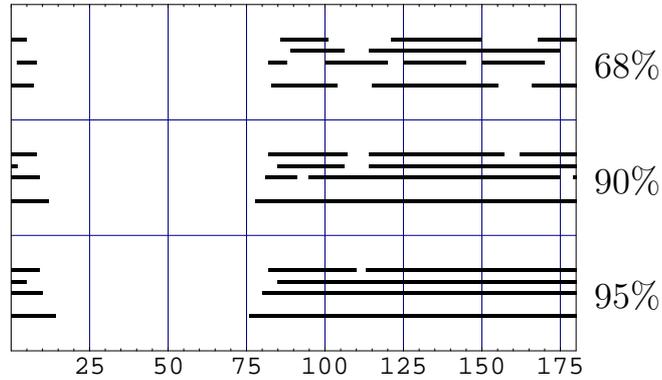,width=0.65\textwidth}
\caption{$\alpha$ regions (the ordering, top to bottom, is in each case: 'PLD' parametrization, '1i' parametrization, Explicit extraction and frequentist analysis).} \label{FIG:AlphaRegions}
\end{center}
\end{figure}

Figure \ref{FIG:Alpha:01} shows the PDF of $\alpha$ in three different cases: the 'PLD' \cite{Pivk:2004hq} and '1i' (\eq{EQ:1iparam:01}) parametrizations, and the explicit extraction (as in \cite{Pivk:2004hq} or \cite{Botella:2005ks}). Corresponding 68\%, 90\% and 95\% probability regions are displayed in table \ref{TAB:AlphaRegions:01}, together with the frequentist 68\%, 90\% and 95\% CL regions (in the following, frequentist calculations are carried with the 'PLD' parametrization). These regions are represented in figure \ref{FIG:AlphaRegions}. Despite some small differences in the 68\% regions, somehow expectable as they are more sensitive to details, the results are consistent, they coincide rather well. $B\to\pi\pi$ data are still too uncertain to really provide important constraints on $\alpha$, the only relevant feature being the exclusion of the $\alpha\sim\pi/4$ region, which could be understood (see section \ref{AP:03:01} in appendix \ref{AP:03}) in terms of the smallness of $B^{00}$. 

\begin{table}[h]
\begin{center}
\begin{tabular}{|c||c|c|c|}
\hline
 & 68\% & 90\% & 95\% \\ \hline\hline
PLD & {\small $[0;5]^\circ \cup [85;101]^\circ\cup$}& {\small $[0;8]^\circ \cup [82;107]^\circ\cup$}& {\small $[0;9]^\circ \cup [82;110]^\circ$}\\ 
 & {\small $ [121;150]^\circ\cup [168;180]^\circ$} & {\small $ [114;157]^\circ\cup [162;180]^\circ$}& {\small $ \cup[113;180]^\circ$}\\\hline

1i &{\small $[95;174]^\circ$}&{\small $[0;1]^\circ \cup [89;180]^\circ$}& {\small $[0;5]^\circ \cup [85;180]^\circ$}\\ \hline

 &{\small $[2;8]^\circ \cup [82;88]^\circ\cup$}&{\small $[0;9]^\circ \cup [81;91]^\circ$}&\\
Explicit  &{\small $[100;120]^\circ \cup [125;145]^\circ\cup$}&{\small $[95;175]^\circ \cup [179;180]^\circ$}&{\small $[0;10]^\circ \cup [80;180]^\circ$}\\ 
 &{\small $[150;170]^\circ$}& &\\ \hline

CL &{\small $[0;7]^\circ \cup [83;104]^\circ$}&{\small $[0;12]^\circ \cup [78;180]^\circ$}& {\small $[0;14]^\circ \cup [76;180]^\circ$}\\
 & {\small $ [115;154]^\circ\cup [166;180]^\circ$} & & \\\hline
\end{tabular}
\caption{$\alpha$ regions within $[0;180^\circ]$.}\label{TAB:AlphaRegions:01}
\end{center}
\end{table}

\clearpage
\newpage

\section*{Conclusions}\label{SEC:Conclusions}

To our knowledge the discrepancies between frequentist and bayesian approaches using the so-called MA and RI parametrizations with \eq{EQ:SMparam} have not been previously understood. We explain that with present experimental uncertainties it is extremely unsecure to introduce the phase $\alpha$ in the $\Delta I=1/2$ piece. To a great extent a spurious PDF of $\alpha$ tends to be generated. The Gronau and London analysis is critically based on the appearance of one weak phase in the $\Delta I=3/2$ piece ($\Cpm=0$). Introducing $\alpha$ in the $\Delta I=1/2$ piece -- or $\Apm$ -- ($\Cpm\neq 0$) brings this ``second'' $\alpha$ to the category of 'not observable' even if one is using a Standard Model inspired parametrization.
This difficulty is operative in the so-called MA and RI parametrizations. The introduction of $\alpha$ in the $\Delta I=1/2$ piece and some QCD-based bounds on the amplitudes allows  -- as done by the UTfit collaboration -- to eliminate the solutions around $\alpha\sim 0,\pi$ inside the SM. The PDF can still be partially contaminated with the spurious $\alpha$ distribution. In $B\to\pi\pi$ it is not dramatic but it could be so in other channels.
This last procedure cannot be applied to an analysis with NP in loops. Therefore, we strongly recommend to use parametrizations where $\alpha$ is just included in the $\Delta I=3/2$ piece. We partially agree with the UTfit collaboration that, in spite of the differences among the frequentist and bayesian methods, both approaches give similar results if one uses parametrizations with a clear physical meaning. In this sense the most relevant result is the exclusion of the region $\bar\alpha\sim 25^\circ-75^\circ$.

\section*{Acknowledgments}\label{SEC:Ack}
This research has been supported by European FEDER, Spanish MEC under grant FPA 2005-01678, \emph{Generalitat Valenciana} under GVACOMP 2007-172, by \emph{Funda\c{c}\~{a}o para a Ci\^{e}ncia e a Tecnologia} (FCT, Portugal) through the projects PDCT/FP/63912/2005, PDCT/FP/63914/2005, CFTP-FCT UNIT 777, and by the Marie Curie RTN MRTN-CT-2006-035505. M.N. acknowledges financial support from FCT. The authors thank J. Bernab\'eu and P. Paradisi for reading the manuscript and useful comments.

\newpage
\appendix
\section{Inputs and numerical methods}\label{AP:01}
Along this work we use the set of experimental measurements \cite{Aubert:2007mj,Aubert:2006fh,Aubert:2005av,Aubert:2004aq,Abe:2006cc,Abe:2005dz,Abe:2004mp,Abe:2004us,Chao:2003ue,Abe:2003yy,Abe:2003ja}, combined by the Heavy Flavour Averaging Group \cite{HFAG}, in table \ref{TAB:experimental}.
\begin{table}[h]
\begin{center}
\begin{tabular}{|c||c||c|}
\hline
$\Bpm_{\pi\pi}$ & $\Boo_{\pi\pi}$ & $\Bpo_{\pi\pi}$\\ \hline
$5.2\pm 0.2$ & $1.31\pm 0.21$ & $5.7\pm 0.4$\\ \hline\hline
$\Cpm_{\pi\pi}$ & $\Spm_{\pi\pi}$ & $\Coo_{\pi\pi}$\\ \hline
$-0.39\pm 0.07$ & $-0.59\pm 0.09$ & $-0.37\pm 0.32$\\ \hline
\end{tabular}
\caption{Experimental results, branching ratios are multiplied by $10^{-6}$.}\label{TAB:experimental}
\end{center}
\end{table}

In terms of $B\to\pi\pi$ amplitudes,
\begin{equation}
 B^{ij}=\tau_{B^{i+j}}\frac{|A^{ij}|^2+|\bar A^{ij}|^2}{2},\quad C^{ij}=\frac{|A^{ij}|^2-|\bar A^{ij}|^2}{|A^{ij}|^2+|\bar A^{ij}|^2},\quad S^{ij}=\frac{2\text{ Im}(\bar A^{ij}{A^{ij}}^\ast)}{|A^{ij}|^2+|\bar A^{ij}|^2}~.\label{EQ:Observables}
\end{equation}
All frequentist CL computations are performed by: (1) minimizing $\chi^2$ with respect to all parameters except the one of interest which is fixed (in this case $\alpha$), (2) computing the corresponding CL through an incomplete $\Gamma$ function.
All bayesian PDFs are computed using especially adapted Markov Chain MonteCarlo techniques.

\section{Experimental results and isospin relations}\label{AP:02}
The isospin relations
\begin{eqnarray}
\Apm+\sqrt 2\Aoo&=&\sqrt 2\Apo~, \notag\\
\ApmB+\sqrt 2\AooB&=&\sqrt 2\ApoB~,\label{EQ:isospin}
\end{eqnarray}
define two triangles in the complex plane whose relative orientation fixes $\alpha$. The sizes of the different sides follow from \eq{EQ:isospin}.

\begin{table}[h]
\begin{center}
\begin{tabular}{|c|c|c||c|c|c|}
\hline
$|\Apm|$ & $\sqrt 2|\Aoo|$ & $\sqrt 2|\Apo|$ & $|\ApmB|$ & $\sqrt 2|\AooB|$ & $\sqrt 2|\ApoB|$\\ \hline
1.441& 1.040& 2.634& 2.176& 1.533& 2.634\\ \hline
\end{tabular}
\caption{Numerical values of the sides of the isospin triangles computed with experimental central values, to be multiplied by $10^{-3}\text{ ps}^{-1/2}$.}\label{TAB:sides}
\end{center}
\end{table}

This allows the reconstruction, up to a number of discrete ambig\"{u}ities - namely up to eight -, of both triangles. Central values of present measurements yield the values of the sides in table \ref{TAB:sides}. One straightforward question is mandatory: do those would-be triangles ``close''? The answer is in the negative because
\begin{eqnarray*}
 |\Apm|+\sqrt 2|\Aoo|=2.481 &{\boldsymbol{\ngtr}}& 2.634=\sqrt 2|\Apo|~,\\
 |\ApmB|+\sqrt 2|\AooB|=3.709 &>& 2.634=\sqrt 2|\ApoB|~.
\end{eqnarray*}
In fact, for those central values, the first triangle \emph{is not} a triangle \cite{Botella:2006zi}. In terms of likelihood, the closest configuration to that situation, the most likely one, is having the first triangle \emph{flat}, a feature which naturally explains the reduced -- by a factor of two, from eight to four -- degeneracy of $\alpha$ ``solutions''. That is, while for old data the almost flatness of this same isospin triangle yielded eight different solutions distributed in four almost-degenerate pairs, those pairs are now degenerate and rather than exact solutions for the central values of the observables they produce best-fitting points.

Consequently, the use of explicit solution constructions requires the rejection of the joint regions of experimental input incompatible with the isospin relations \eqs{EQ:isospin}. For old data, this meant rejecting some 48.2\% of allowed experimental input (weighting each observable with a gaussian with mean and standard deviation given by the corresponding central value and uncertainty), for the new data set this rejection rate is 70.9\%. In the bayesian and frequentist treatments the isospin relations are assumed valid and all the subsequent analyses are ``normalized'' to that assumption.

\section{Removing $\boldsymbol{B\to\pi^0\pi^0}$ information}\label{AP:03}

\subsection{Explicit extraction of $\boldsymbol{\alpha}$}\label{AP:03:01}

\begin{figure}[h]
\begin{center}
\subfigure[$\alpha$ PDF\label{FIG:ExplicitNo00:a}]{\epsfig{file=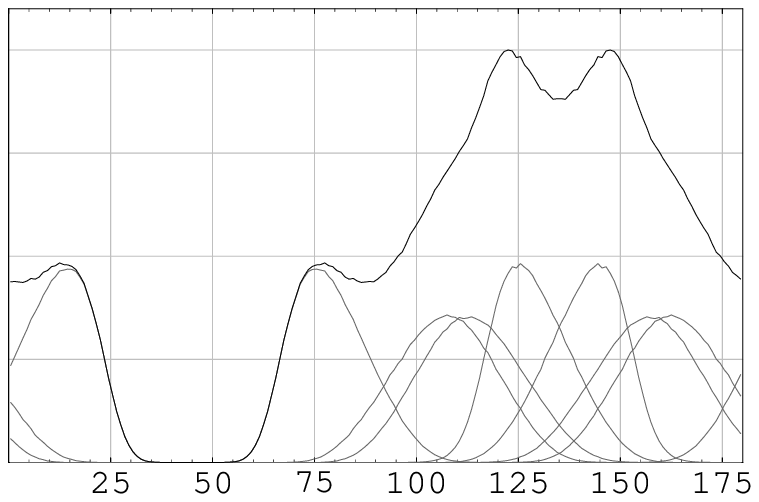,width=0.45\textwidth}}\qquad
\subfigure[$\alpha$ PDF\label{FIG:ExplicitNo00:b}]{\epsfig{file=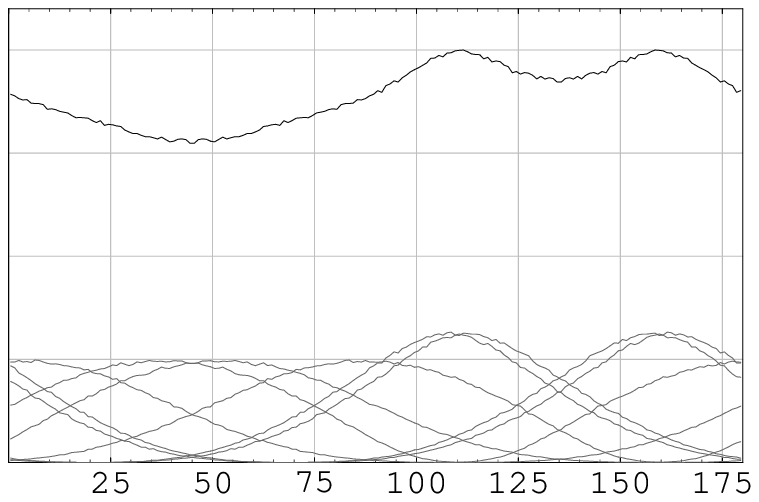,width=0.45\textwidth}}
\caption{Explicit extraction without $B\to\pi^0\pi^0$; lighter curves correspond to the different individual contributions related by the discrete ambig\"{u}ities.} \label{FIG:ExplicitNo00}
\end{center}
\end{figure}

This appendix is devoted to some complementary results extending what is presented in section \ref{sSEC:02:01}. The first issue we will address is the explicit extraction\footnote{Beside the explicit formula for $\alpha$ in terms of the available observables presented in reference \cite{Pivk:2004hq} we also make use of the extraction of $\alpha$ explained in \cite{Botella:2005ks}; the results are completely equivalent, however the later does not make any use of a particular parametrization of the amplitudes and is easily interpreted in terms of the isospin construction.} of $\alpha$ when $B\to\pi^0\pi^0$ information is removed, that is, no knowledge of $\Boo$ and $\Coo$. The explicit extraction of $\alpha$ assumes the isospin relations in \eqs{EQ:isospin} so to start with, the \emph{ignorance} on $\Boo$ is not "just plain ignorance" (whatever this could stand for) as it will operatively mean that for any experimental set of results $\{\Bpm,\Bpo,\Cpm,\Spm\}$, $\Boo$ and $\Coo$ should be such that both would-be isospin triangles \emph{are} in fact isospin triangles. $\Coo$ is obviously restricted to be in the range $[-1;1]$; what about $\Boo$? One could argue that if there is no information on $B\to\pi^0\pi^0$ it should be smaller than a given bound or one can just let it be as large as allowed by other data and isospin constraints. This rather trivial fact is apparently at the origin of the discrepancy in the results presented in references \cite{Charles:2006vd,Bona:2007qt} for the explicit extraction of $\alpha$ ``without'' $B\to\pi^0\pi^0$ information: figure \ref{FIG:ExplicitNo00} shows two PDFs of $\alpha$. They are obtained by generating known experimental sets $\{\Bpm,\Bpo,\Cpm,\Spm\}$ according to gaussian distributions with central values and standard deviations given by the quoted measurements and uncertainties ($\Cpm$ and $\Spm$ are also restricted to be within $[-1;1]$), then $\Coo$ and $\Boo$ are generated through flat distributions, $\Coo$ in the range $[-1;1]$ and $\Boo$ in a range $[0;\Boo_{Max}]$. Sets $\{\Bpm,\Bpo,\Cpm,\Spm,\Boo,\Coo\}$ which fulfill the isospin relations \eqs{EQ:isospin} are retained and used to extract $\alpha$. The PDFs of $\alpha$ represented in figure \ref{FIG:ExplicitNo00} only differ in the value of $\Boo_{Max}$, Fig. \ref{FIG:ExplicitNo00:a} was obtained with $\Boo_{Max}$ equal to \emph{two} times the present measurement while Fig. \ref{FIG:ExplicitNo00:b} was obtained with $\Boo_{Max}$ equal to \emph{twenty} times the present measurement. On the one hand, the PDF in figure \ref{FIG:ExplicitNo00:a} coincides with the one presented in figure 4 'ES' of reference \cite{Charles:2006vd}; on the other hand the PDF in figure \ref{FIG:ExplicitNo00:b} agrees, more or less, with figure 4 of reference \cite{Bona:2007qt}. It is now clear that the difference among both may be just due to the numerical procedure. Figure \ref{FIG:ExplicitNo00:b} shows that the removal of $B\to\pi^0\pi^0$ information leads to a loss of knowledge on $\alpha$. Ironically, there is a lesson in this example: numerics apart, the smallness of $\Boo$ is responsible for the exclusion of values $\alpha\sim \pi/4$.

\subsection{Parametrizations}\label{AP:03:02}
\begin{figure}[h]
\begin{center}
\subfigure[$\alpha$ PDF, PLD parametrization\label{FIG:No00:1a}]{\epsfig{file=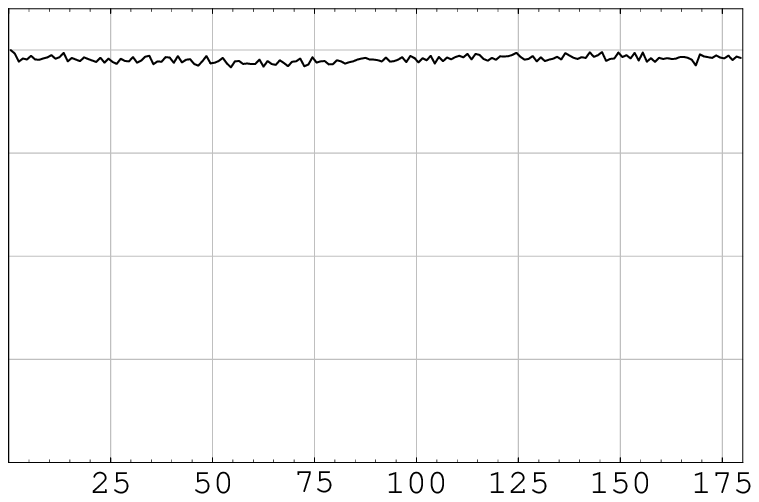,width=0.45\textwidth}}\qquad
\subfigure[$\alpha$ PDF, 1i parametrization\label{FIG:No00:1b}]{\epsfig{file=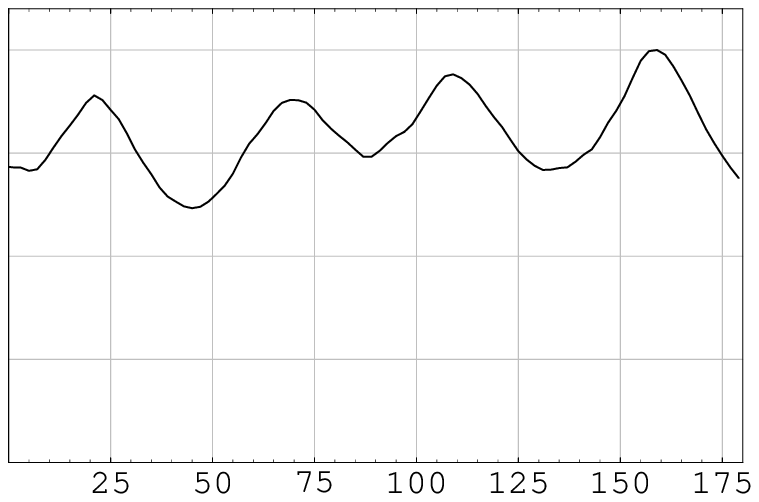,width=0.45\textwidth}}
\caption{Extraction without $B\to\pi^0\pi^0$.} \label{FIG:No00:1}
\end{center}
\end{figure}

To complete the picture we now proceed to repeat the extraction of $\alpha$ when $B\to\pi^0\pi^0$ information is removed in several parametrizations. We will make use of the 'PLD' parametrization \cite{Pivk:2004hq}, of the '1i' parametrization with fixed weak phases in $\{\Apm,\ApmB\}$ (\eq{EQ:1iparam:01}) and, finally, of the parametrization in \eq{EQ:AlphaPrime:Parametrization} but in this case, apart from $\alpha$ and $\alpha^\prime$, instead of moduli and phases we will use real and imaginary parts of $\Tpm$, $\Pe$ and $\Too$ (the RI parametrization in reference \cite{Charles:2006vd}). The PDFs of $\alpha$ obtained for the first two parametrizations are shown in figure \ref{FIG:No00:1}, they are eloquent: no knowledge on $\alpha$.

\begin{figure}[h]
\begin{center}
\subfigure[$\alpha$ PDF\label{FIG:No00:2a}]{\epsfig{file=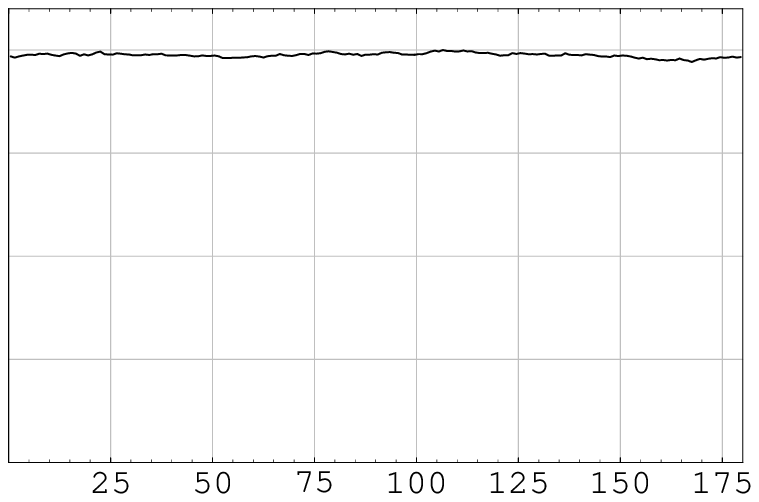,width=0.3\textwidth}}\quad
\subfigure[$\alpha^\prime$ PDF\label{FIG:No00:2b}]{\epsfig{file=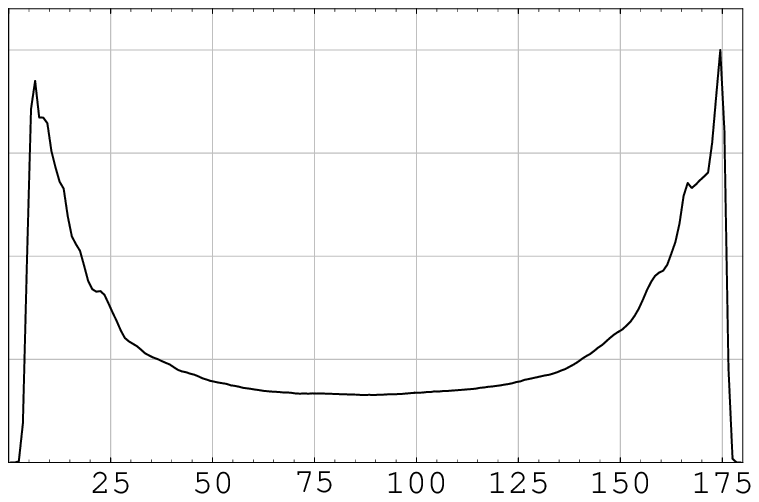,width=0.3\textwidth}}\quad
\subfigure[$\alpha=\alpha^\prime$ PDF\label{FIG:No00:2c}]{\epsfig{file=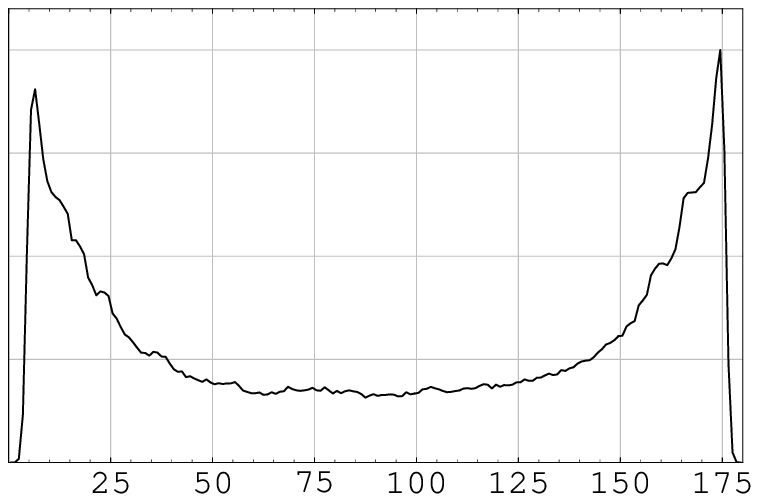,width=0.3\textwidth}}
\caption{Extraction without $B\to\pi^0\pi^0$, RI parametrization.} \label{FIG:No00:2}
\end{center}
\end{figure}

For the RI parametrization we show the PDFs of $\alpha$, $\alpha^\prime$ and the one obtained by setting $\alpha=\alpha^\prime$ in figure \ref{FIG:No00:2}. Once again it is clear that there is no information on $\alpha$ and that inappropriately insisting on including it in $\{\Apm,\ApmB\}$ produces the senseless result of figure \ref{FIG:No00:2c}.

The conclusion of this appendix is straightforward: just dealing with a reduced scenario in which $B\to\pi^0\pi^0$ information is removed, a proper understanding of the subtleties involved in the parametrization of $B\to\pi\pi$ amplitudes avoids peculiar results as for instance the 'MA' and 'RI' ones included in figure 4 of reference \cite{Charles:2006vd}. We have shown here that starting with a flat prior for $\alpha$ consistently gives highly non-informative posteriors in several sensible parametrizations.

\section{Using the RI parametrization}\label{AP:RI}

\begin{figure}[h]
\begin{center}
\subfigure[$\alpha$ PDF\label{FIG:RIa}]{\epsfig{file=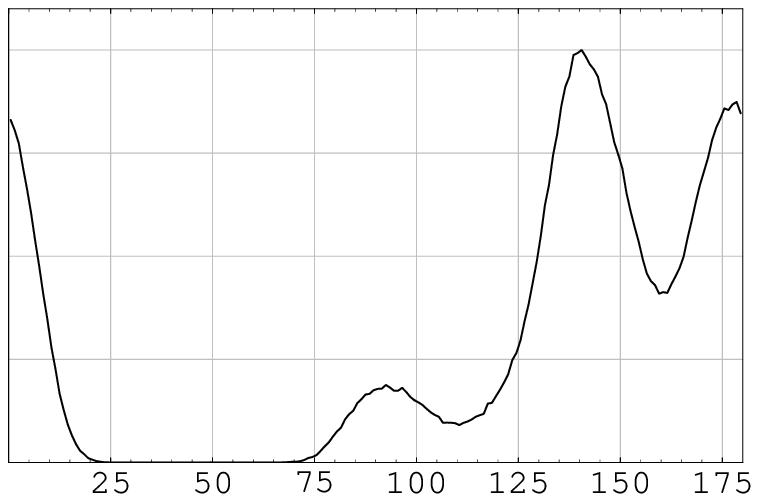,width=0.3\textwidth}}\quad
\subfigure[$\alpha^\prime$ PDF\label{FIG:RIb}]{\epsfig{file=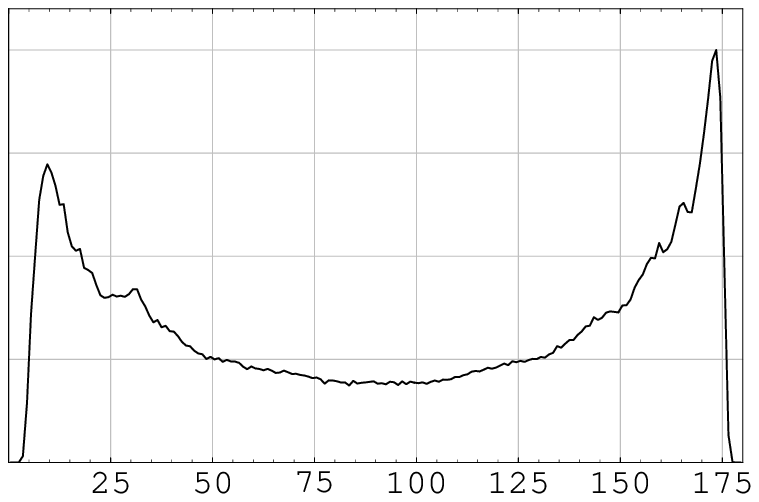,width=0.3\textwidth}}\quad
\subfigure[$\alpha=\alpha^\prime$ PDF\label{FIG:RIc}]{\epsfig{file=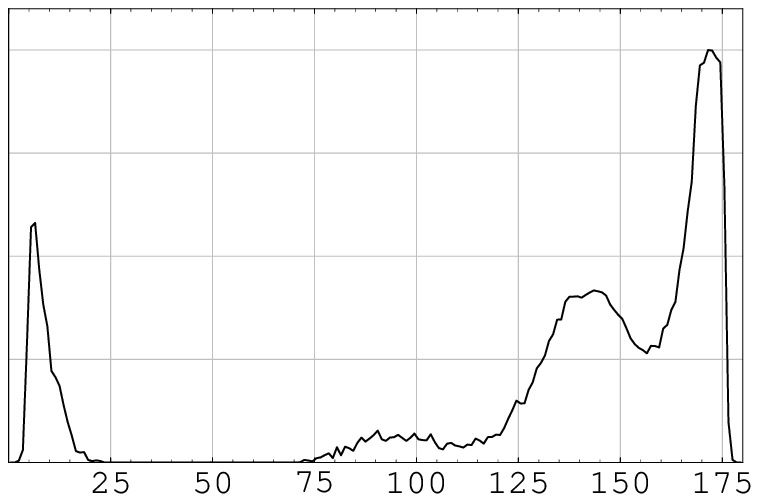,width=0.3\textwidth}}
\caption{$\alpha$ extraction, RI parametrization.} \label{FIG:RI}
\end{center}
\end{figure}

In section \ref{sSEC:02:02} we used the parametrization in \eq{EQ:AlphaPrime:Parametrization} to obtain figure \ref{FIG:alpha:alphaprime:02} with flat $|\Tpm|$, $|\Pe|$, $|\Too|$, $\arg(\Pe)$, $\arg(\Too)$, $\alpha$ and $\alpha^\prime$ priors. For completness we also show -- figure \ref{FIG:RI} -- the PDFs of $\alpha$, $\alpha^\prime$ and $\alpha=\alpha^\prime$ in case one uses flat $|\Tpm|$, $\re{\Pe}$, $\im{\Pe}$, $\re{\Too}$, $\im{\Too}$, $\alpha$ and $\alpha^\prime$ priors. Beside the effect of the spurious $\alpha^\prime$ in the PDF of $\alpha=\alpha^\prime$, we can also appreciate the influence of the change in the priors: the integration domain is the same as in figure \ref{FIG:alpha:alphaprime:02} but the integration measure is now different. The main effect is the relative enhancement of the contributions from regions with large parameters, including the contributions from the $\alpha^\prime\to 0$ driven region.

\newpage
\section{One short statistical comment}\label{AP:04}
\begin{figure}[hb]
\begin{center}
\subfigure[Joint $(\delta,\alpha)$ PDF\label{FIG:Joint:1a}]{\epsfig{file=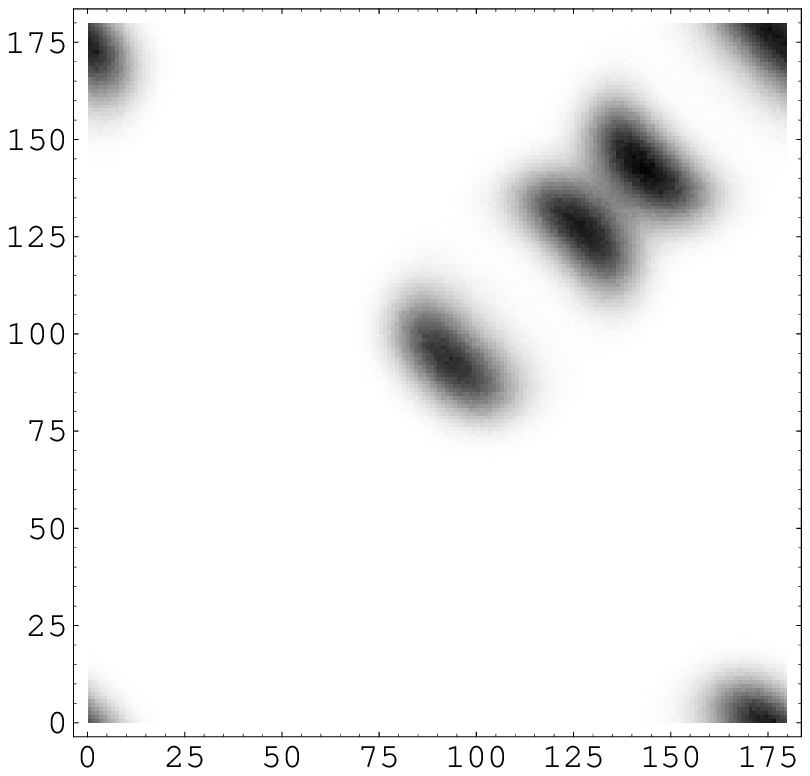,width=0.4\textwidth}}\quad
\subfigure[Joint $(\alpha_{eff},\alpha)$ PDF\label{FIG:Joint:1b}]{\epsfig{file=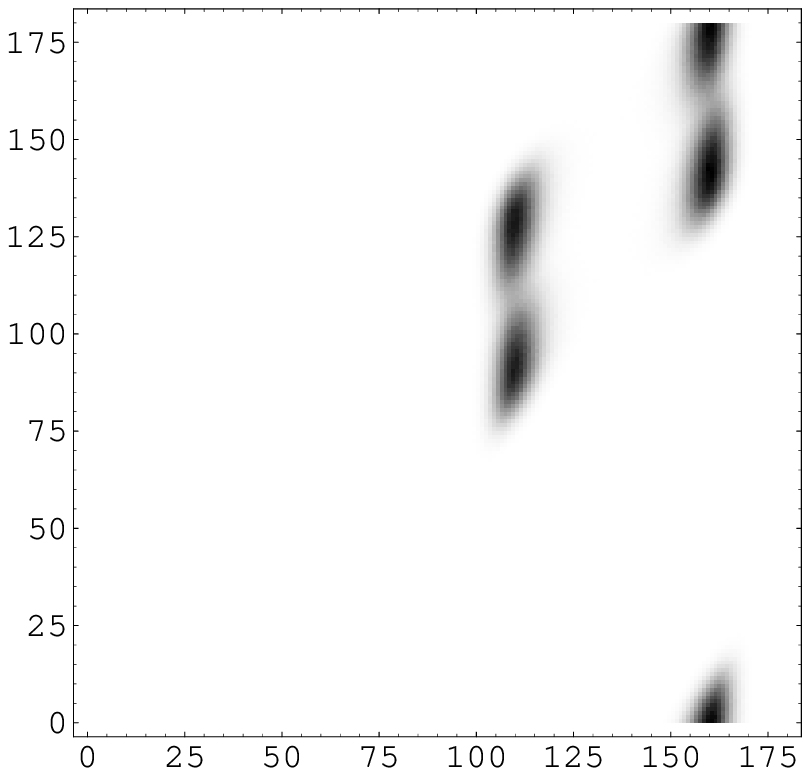,width=0.4\textwidth}}
\caption{Joint PDFs obtained with the 'PLD' parametrization.} \label{FIG:Joint}
\end{center}
\end{figure}

Leaving completely aside philosophical aspects of probability, both frequentist and bayesian approaches start with a common likelihood function. Each approach reduces the information provided by the likelihood function in a different manner. Consequently, they do not yield strictly coincident results:
\begin{itemize}
 \item Bayesian posteriors obviously depend on the priors, for example the allowed ranges or the shape. As we have seen, we obtain different posteriors with different priors. However, as long as one is using sensible parametrizations and reasonable priors, we end up finding rather compatible results.
 \item Frequentist CL curves do depend on the parametrization, to be precise, they depend on the allowed ranges for the parameters; once sensible parametrizations and adequate ranges are used, CL curves obtained with them are identical. The $\alpha\to 0$ limit in the SM inspired parametrization of \eq{EQ:SMparam} illustrates this issue.
\end{itemize}
Beside those well known issues, we may find troublesome that:
\begin{enumerate}
 \item Most probable values in the bayesian PDFs do not coincide with the analytical solutions for $\alpha$.
 \item Intimately related to this aspect, bayesian PDFs seem unable to distinguish among degenerate solutions.
\end{enumerate}
We remind that these statements concern one dimensional PDFs of $\alpha$. Frequentist one dimensional CL curves distinguish $\alpha$ solutions because they are obtained through best fitting points for fixed $\alpha$. Bayesian PDFs do not distinguish them as the uncertainties produce distributions for the degenerate solutions which overlap and add up in the complete PDF. One can still have a hint of the proximity of different solutions from this kind of overlap, but this is not the point here. For reduced experimental uncertainties, bayesian PDFs would not overlap and would distinguish among those different solutions. This could be sufficient to think that, \emph{per se}, there is no discriminating advantage in using one or the other approach. With present uncertainties, bayesian analyses seem incapable of pinning down the right location of the solutions in $\alpha$ and telling us something about their degeneracy. It is not a fundamental problem of bayesian methods as reduced uncertainties would overcome these ``difficulties''. If it is not a fundamental problem, could we somehow overcome these ``difficulties'' with present uncertainties? The answer is in the positive as the problem only arises because we are insisting in the reduction of the available experimental information to obtain one-dimensional PDFs of $\alpha$; let us take a look to the joint PDFs in figure \ref{FIG:Joint}. These are the joint PDFs of $(\delta,\alpha)$ and $(\alpha_{eff},\alpha)$ obtained with the 'PLD' parametrization. They are quite illustrative, one can see the different solutions in $\alpha$ concentrated around the values of $\alpha$ dictated by the analytical expectations. The pretended fundamental drawbacks of bayesian methods to adequately place and distinguish the solutions are just a consequence of pushing too far, for the present level of experimental uncertainty in the results, the statistical ``reduction of information process''. A simultaneous look to both frequentist and bayesian results will not put an end to the statistical discrepancies, notwithstanding it will be very helpful to understand the physical results we are interested in. Both approaches are ``information reduction processes'' and strictly sticking to one and deprecating the other may not be the wiser strategy. 

\end{document}